\documentclass[a4paper,fleqn]{cas-sc}

\usepackage[square,numbers,compress,sort]{natbib}
\makeatletter
\renewcommand\@biblabel[1]{#1.}
\makeatother
\usepackage{csquotes}
\usepackage{xcolor}
\usepackage[colorinlistoftodos,prependcaption]{todonotes}
\usepackage[inline]{enumitem}
\usepackage{xspace}
\usepackage{multicol}
\usepackage{color}
\usepackage{acronym} 
\usepackage{subcaption}
\usepackage{longtable}
\usepackage{afterpage}
\usepackage{csquotes}
\usepackage{amssymb}%
\usepackage{pifont}
\newcommand{\cmark}{\ding{51}}%
\newcommand{\xmark}{\text{\ding{55}}}
\usepackage[commandnameprefix=ifneeded]{changes}

\def\tsc#1{\csdef{#1}{\textsc{\lowercase{#1}}\xspace}}
\tsc{WGM}
\tsc{QE}
\tsc{EP}
\tsc{PMS}
\tsc{BEC}
\tsc{DE}

\usepackage{float}
\usepackage[usenames,dvipsnames]{pstricks}
\usepackage{pstricks-add}
 \usepackage{epsfig}
 \usepackage{pst-grad} 
 \usepackage{pst-plot} 
 \usepackage[space]{grffile} 
 \usepackage{etoolbox} 
 \makeatletter 
 \patchcmd\Gread@eps{\@inputcheck#1 }{\@inputcheck"#1"\relax}{}{}
 \makeatother
\usepackage{verbatim}
\usepackage{multicol}
\usepackage[ruled,lined,noresetcount ]{algorithm2e}
\usepackage{algpseudocode}
\usepackage{mathtools}

\begin{document}
\let\WriteBookmarks\relax
\def\floatpagepagefraction{1}
\def\textpagefraction{.001}

\shorttitle{{Deep Learning for Steganalysis of Diverse Data Types: A review of methods, taxonomy, challenges and future directions}}

\shortauthors{Dr. H Kheddar et~al.}

\title [mode = title]{{Deep Learning for Steganalysis of Diverse Data Types: A review of methods, taxonomy, challenges and future directions}}

 
\vskip2mm

\author[1]{Hamza Kheddar\corref{cor1}}
[orcid=0000-0002-9532-2453]
\ead{kheddar.hamza@univ-medea.dz}
\cormark[1]
\credit{Conceptualization; Methodology; Data Curation; Resources; Investigation; Visualization;  Writing original draft; Writing, review, and editing}

\author[2]{Mustapha Hemis}
[orcid=0000-0002-6353-0215]
\ead{hemismustapha@yahoo.fr }
\credit{Conceptualization; Methodology; Resources; Investigation; Writing original draft; Writing, review, and editing}

\author[3]{Yassine Himeur}
[orcid=0000-0001-8904-5587]
\ead{yhimeur@ud.ac.ae}
\credit{Conceptualization; Methodology; Resources; Investigation; Writing original draft; Writing, review, and editing}

\author[4]{David Meg\'{i}as} 
[orcid=0000-0002-0507-7731]
\ead{dmegias@uoc.edu}
\credit{Conceptualization; Methodology; Resources; Visualization; Investigation;   review, and editing}

\author[5,6]{Abbes Amira}
[orcid=0000-0003-1652-0492]
\ead{aamira@sharjah.ac.ae,abbes.amira@dmu.ac.uk}
\credit{Methodology; Resources; Visualization; Investigation; Writing – review and editing, Supervision}

\address[1]{LSEA Laboratory, Department of Electrical Engineering, University of Medea, 26000, Algeria}

\address[2]{LCPTS Laboratory, University of Sciences and Technology Houari Boumediene (USTHB), P.O. Box 32, El-Alia, Bab-Ezzouar, Algiers 16111, Algeria.}

\address[3]{College of Engineering and Information Technology, University of Dubai, Dubai, UAE}

\address[4]{Internet Interdisciplinary Institute (IN3), Universitat Oberta de Catalunya (UOC), CYBERCAT-Center for Cybersecurity Research of Catalaonia, Rambla del Poblenou, 154, 08018 Barcelona, Spain.}

\address[5]{Department of Computer Science, University of Sharjah, UAE}
\address[6]{Institute of Artificial Intelligence, De Montfort University, Leicester, United Kingdom}

\tnotetext[1]{The first is the corresponding author.}

\begin{abstract}
Steganography and steganalysis are two interrelated aspects of the field of information security. Steganography seeks to conceal communications, whereas steganalysis aims to discover or, if possible, recover the data they contain. These two areas have garnered significant interest, especially among law enforcement agencies. Cybercriminals and even terrorists often employ steganography to avoid detection while in possession of incriminating evidence, even when that evidence is encrypted, since cryptography is prohibited or restricted in many countries. Therefore, a deep understanding of cutting-edge techniques for uncovering concealed information is essential in exposing illegal activities. Over the last few years, a number of strong and reliable steganography and steganalysis techniques have been introduced in the literature.  This review paper provides a comprehensive overview of deep learning-based steganalysis techniques used to detect hidden information within digital media. The paper covers all types of cover in steganalysis, including image, audio, and video, and discusses the most commonly used deep learning techniques. In addition, the paper explores the use of more advanced deep learning techniques, such as \ac{DTL} and \ac{DRL}, to enhance the performance of steganalysis systems. The paper provides a systematic review of recent research in the field, including data sets and evaluation metrics used in recent studies. It also presents a detailed analysis of DTL-based steganalysis approaches and their performance on different data sets. The review concludes with a discussion on the current state of deep learning-based steganalysis, challenges, and future research directions. 
\end{abstract}




\maketitle


\begin{table}[]
    \centering
{\small \section*{Acronyms and Abbreviations}}
\begin{multicols}{3}
\footnotesize
\begin{acronym}[AWGN] 
\acro{AAC}{advanced audio coding}
\acro{AE}{auto encoder}
\acro{AMR}{adaptive multi-rate}
\acro{AUC}{area under the curve}
\acro{AWGN}{additive white gaussian noise}
\acro{BERT}{bidirectional encoder representations from transformers }
\acro{BN}{batch normalization}
\acro{bpp}{bit per pixel}
\acro{BPTT}{back propagation through time}
\acro{CBAM}{convolution block attention module}
\acro{CD}{contrastive divergence}
\acro{CMV}{candidate motion vector}
\acro{CNN}{convolution neural network}
\acro{CSA}{channel-spatial attention}
\acro{CSM}{cover source mismatch}
\acro{DAE}{denoising auto-encoder}
\acro{DBN}{deep belief network}
\acro{DCNN}{densely connected convolutional neural network}
\acro{DCT/DST}{discrete Cosine/Sine transform}
\acro{DCT}{discrete cosine transform}
\acro{DER}{detection error rate}
\acro{DL}{deep learning}
\acro{DNA}{deoxyribonucleic acid}
\acro{DNN}{deep neural network}
\acro{DNNITS}{deep neural network-based invisible text steganalysis}
\acro{DR}{detection rate}
\acro{DRL}{deep reinforcement learning}
\acro{DRN}{deep residual network}
\acro{DT}{destruction rate}
\acro{DTL}{deep transfer learning}
\acro{EA}{evolutionary algorithm}
\acro{ECA}{efficient channel Attention}
\acro{EECS}{equal length entropy codes substitution}
\acro{FAR}{false alarm rate}
\acro{FCB}{fixed codebook}
\acro{FFT}{fast Fourier transform}
\acro{FLOP}{floating-point operations}
\acro{FN}{false negative}
\acro{FP}{false positive}
\acro{FPR}{false positive rate}
\acro{GMM}{Gaussian mixture model}
\acro{GNCNN}{Gaussian-neuron convolutional neural network}
\acro{GNN}{graph neural network}
\acro{GPU}{graphics processing unit}
\acro{GRL}{gradient reversal layer}
\acro{GRU}{gated recurrent unit}
\acro{HAS}{human auditory system}
\acro{HEVC}{high efficiency video coding}
\acro{HGD}{high-level representation guided denoiser}
\acro{HPF}{high-pass filtering}
\acro{IoT}{internet of things}
\acro{JPEG}{joint photographic experts group}
\acro{LSB}{least significant bits}
\acro{LSTM}{long short-term memory network}
\acro{MAE}{mean absolute error}
\acro{MFCC}{mel-frequency cepstral coefficients}
\acro{MFRNet}{multi-frequency residual deep convolutional neural network}
\acro{ML}{machine learning}
\acro{MLP}{multilayer perceptron}
\acro{MP3}{MPEG-1 audio layer 3}
\acro{MSE}{mean squared error}
\acro{MV}{Motion vector}
\acro{NAS}{neural architecture search}
\acro{NN}{neural network}
\acro{NSGA-III}{nondominated sorting genetic algorithm-III}
\acro{PSNR}{peak signal-to-noise ratio}
\acro{PU}{prediction unit}
\acro{QIM}{quantization index modulation}
\acro{QMDCT}{quantified modified discrete cosine transform}
\acro{RBM}{restricted Boltzmann machines}
\acro{ReLU}{rectified linear unit }
\acro{ResNet}{residual neural network}
\acro{RMSProp}{root mean squared propagation}
\acro{RNN}{recurrent neural network}
\acro{ROC}{receiver operating characteristic}
\acro{SCA}{selection-channel aware}
\acro{SDA}{subdomain adaptation}
\acro{SGD}{stochastic gradient descent}
\acro{SOTA}{state-of-the-art}
\acro{SPP}{spatial pyramid pooling}
\acro{SRM}{spatial rich model}
\acro{SRM-EC}{spatial rich models with ensemble classifier}
\acro{SSIM}{Structural similarity index}
\acro{STC}{syndrome-trellis code}
\acro{SVM}{support vector machine}
\acro{TanH}{hyperbolic tangent}
\acro{TLU}{truncated linear unit}
\acro{TN}{true negative}
\acro{TP}{true positive}
\acro{TPR}{true positive rate}
\acro{VoIP}{voice over IP}
\acro{VSOFD}{video surveillance object forgery detection}
\end{acronym}

\end{multicols}
\end{table}

\section{Introduction}
\label{sec1}

In recent years, technology has rapidly advanced, leading to the widespread use of multimedia for data transfer, particularly in the realm of the \ac{IoT} \cite{sayed2021intelligent}. However, the transfer of data often occurs over insecure network channels, with the Internet being a popular medium for the exchange of digital media among individuals, private companies, institutions, and governments \cite{himeur2022latest}. While there are many benefits associated with this method of data exchange, one significant drawback is the risk to data privacy and security. The existence of readily available tools capable of exploiting vulnerabilities in data transfer channels has made malicious threats, eavesdropping, and other subversive activities a real possibility \cite{himeur2022blockchain}. 

The primary solution to the security risks associated with data transfer is to use data encryption, including symmetric and asymmetric key cryptography, whereby the data is transformed into a cipher text using an encryption key \cite{rathore2022novel,himeur2018robust}. The receiver can then decrypt the message using a decryption key to obtain the original plain text. While data encryption is an effective way to protect sensitive information, the encrypted data is often unreadable and suspicious in appearance, which can lead to further scrutiny \cite{jing2023fpga,20032016performance}. As a result, another research topic called steganography emerged, which involves hiding data in a way that is statistically undetectable and, also, invisible to the human eye for image \cite{kadhim2019comprehensive}, video \cite{liu2019video}, and text carriers \cite{alanazi2022inclusion}, and imperceptible to the \ac{HAS} for audio carriers, in low \cite{kheddar2019pitch,kheddar2018fourier}, and high \cite{kheddar2022high,kheddar2022speech} encoding rates. This approach allows for the secure transfer of data without arousing suspicions or attracting unwanted attention.

Steganalysis, on the other hand, is the art of detecting secret messages embedded in digital media using steganography. Steganalysis aims to identify suspicious information, determine the occurrence of the message, and, if possible, recover the message. Steganalysis and steganography advancements have increased standards for both. 
Steganalysis could be used to detect and extract hidden information in digital media to reduce the harm caused by those who use steganography tools for malicious purposes. Its primary objective is to estimate whether there is secret information in the media being tested by identifying the steganography type and, in a second phase, extracting the hidden information content. Image steganalysis is the most widely known technique, in which features are extracted through feature design, and different classifiers can be trained to make the technology more adaptable (Figure \ref{fig1}).  Real-time communication requires examining information suspected of having hidden information in media such as images, audio, and video as quickly as possible, but detecting hidden information in real-time is a challenge due to various factors such as bandwidth, delay, and throughput. Combining deep learning with steganalysis can help achieve positive results by automatically extracting the complex and effective messages hidden in the data, which takes less time and resources and makes up for the shortcomings of traditional image steganalysis. 

This review aims to help researchers achieve better outcomes in steganalysis by studying current methodologies and understanding future challenges in digital media steganalysis. The survey provides a summary of different types of algorithms for digital image, video, speech, and text steganalysis that have been developed using classical \ac{ML} and \ac{DL} technologies, with a focus on algorithms for temporal, spatial, and transform domains. Typically, the use of several \ac{DL} algorithms in steganalysis is investigated, such as \ac{DNN}, \ac{CNN}, \ac{RNN}, \ac{LSTM}, autoencoders, \ac{RBM}, \ac{DBN} and \ac{GNN}
.
The survey also begins by providing an overview of research methodology in Section \ref{sec2}, and Section \ref{sec3} introduces a background for both steganalysis and deep learning algorithms. Section \ref{sec4} summarizes most of the carrier-type-based deep steganalysis existing approaches. Section \ref{sec5} discussed existing techniques-type-based deep steganalysis existing approaches. Research gaps and future directions are discussed in Section \ref{sec6}. Finally, Section \ref{sec7} concludes this literature review paper.

\subsection{Background}
\subsubsection{Definition}
\begin{figure}[h!]
\centering
\includegraphics[scale=0.88]{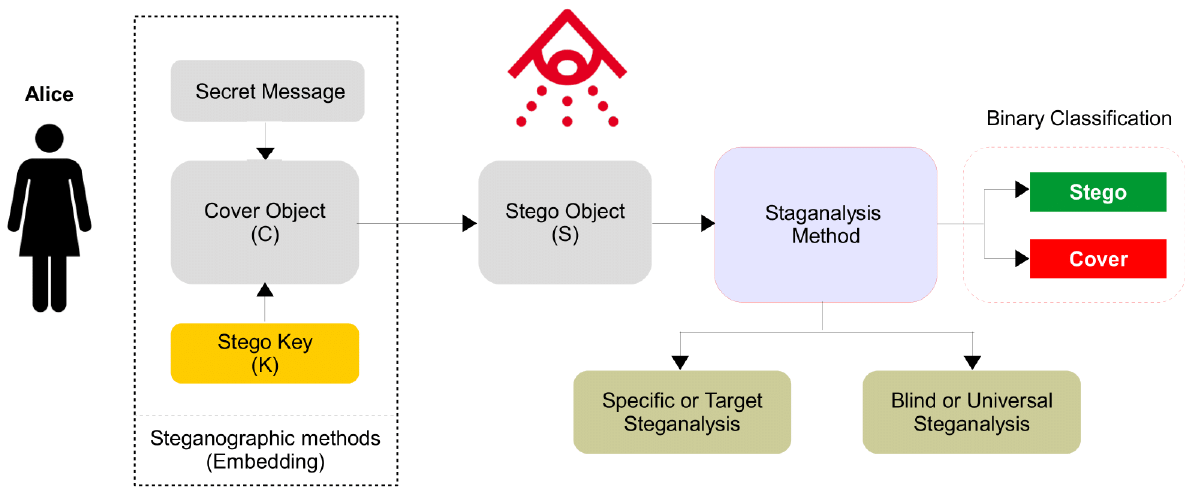}
\caption{Steganalysis process.}
\label{fig1}
\end{figure}

In contrast to steganography, which is used for hiding information within seemingly harmless media, steganalysis is the technology that attempts to defeat steganography--by detecting the hidden information and, in a second stage, if possible, extracting or destroying it.
Figure \ref{fig1} depicts the steganalysis process limited to the detection of the presence of a steganographic message. Additionally, unlike cryptanalysis, which deals with encrypted data, steganalysis entails going through a collection of multimedia data (i.e., audio, image, video, etc.), frequently with no prior understanding of the data that contain a payload. Statistical analysis is employed by steganalysists to reduce the pool of suspect data.

\subsubsection{Mathematical formulation}
The stego data $S$ is often represented by an \ac{AWGN} model commonly used in information theory to mimic the effects of unpredictable natural processes. The \enquote{additive} term refers to the model being added to any noise that may naturally occur in information systems. The \enquote{white} term indicates that the model has uniform power across the frequency band. Finally, the \enquote{Gaussian} term refers to its normal distribution with zero mean value in the time domain. From a statistical point of view, the steganograms can be considered as random variables with the following probability density function \cite{kheddar2019pitch}:

\begin{equation}
\label{eq:gauss}
\mathrm{pdf}(S)=\frac{1}{ \sqrt{2\pi}\sigma}exp\left(-\frac{S^2}{2\sigma^2}\right)
\end{equation}

\begin{figure}[h!]
    \centering
    \includegraphics[scale=0.8]{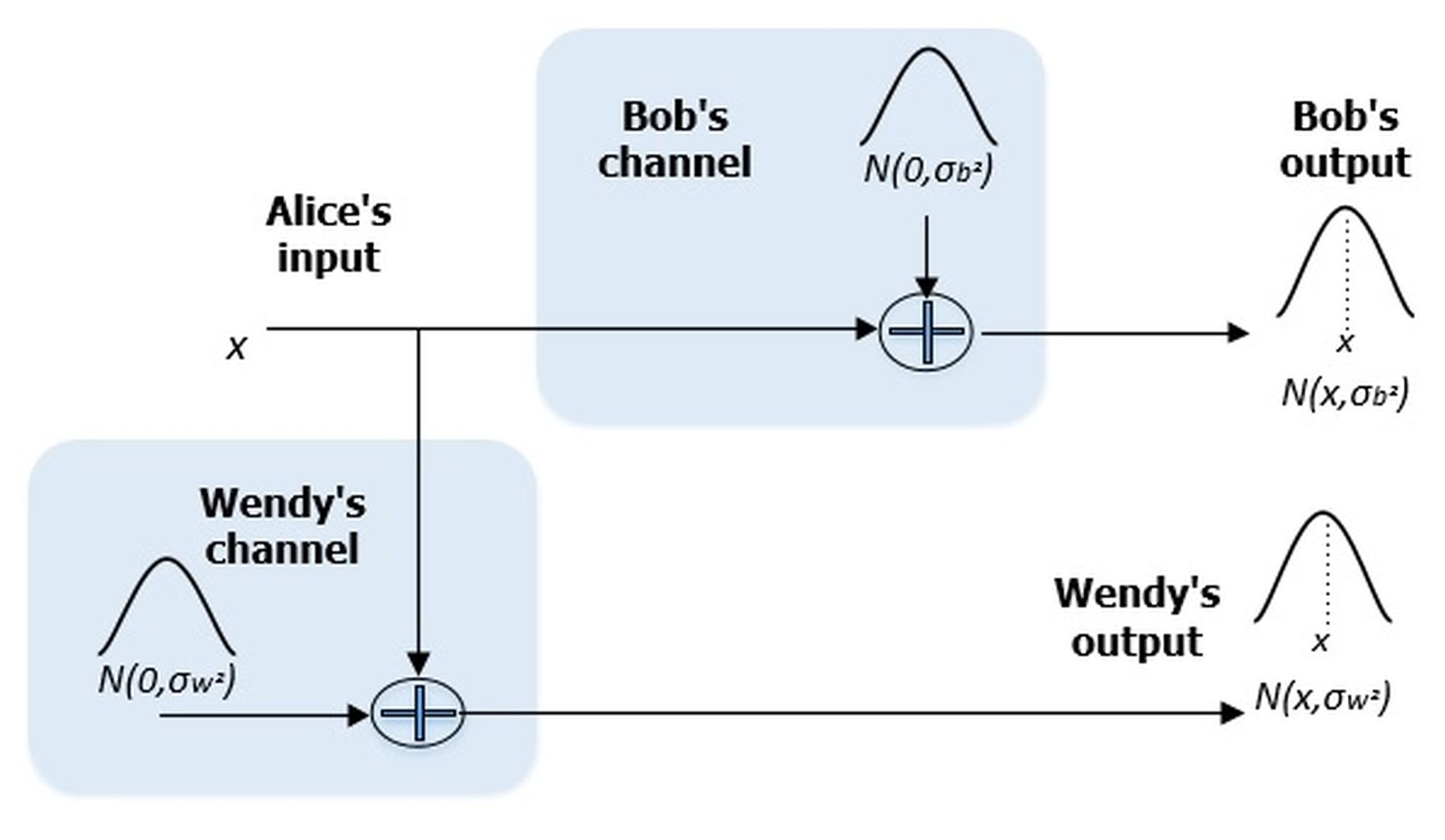}
        \caption{ Interpretation of steganograms in \ac{AWGN} channel.}
    \label{awgn}
\end{figure}

The assumption of an \ac{AWGN} channel model, in which the signal sequence is corrupted by the addition of independent and identically distributed zero-mean Gaussian random variables with variance $\sigma^2$,  is the standard model for a communication channel. We assume that the channels from Alice to Bob and to Wendy are subject to \ac{AWGN} with respective variances $\sigma_b^2>0$ and $\sigma_w^2>0$, as depicted in Figure \ref{awgn}. The steganograms $\xi$ are random variables that follow a normal distribution with  expectation  $E(\xi) = 0$ and variance  $D(\xi) = \sigma^2$.

By using \ac{DL} techniques, the steganalysis algorithms can identify the presence of hidden information in the stego signal ($S$), which differs from the statistical properties of the original signal ($x$). DL-based steganalysis techniques aim to detect the presence of hidden information with high accuracy while minimizing the \ac{FP} and the \ac{FN} rates.

\subsection{Motivation and contribution}

The field of steganalysis-based \ac{DL} has seen significant advancements in recent years with the development of numerous schemes for image, video, text, and speech data. As the prevalence of steganography increases, there is a growing need for efficient and effective steganalysis techniques. \ac{DL} plays a crucial role in steganalysis, the art of detecting hidden information in digital media. It offers numerous benefits, including improved detection accuracy due to its ability to learn complex patterns and features automatically. \ac{DL} models can adapt to diverse data types like images, audio, and video, making them versatile. They reduce the need for manual feature engineering, can handle countermeasures, and scale effectively for large data sets. Some models are optimized for real-time processing, making them suitable for time-sensitive applications. Continuous advancements in \ac{DL} contribute to the ongoing improvement of steganalysis systems. However, challenges such as computational requirements and adversarial techniques must be addressed to maintain effectiveness in this evolving field. A comprehensive survey of proposed schemes in this area is necessary to provide an overview of the \ac{SOTA} approaches and to identify areas for future research.

The contribution of this review paper is to provide an extensive survey of steganalysis-based \ac{DL} techniques across various data types. According to the statistics gathered from the Scopus database, a comparison of existing reviews in the field is summarized in Table \ref{tab:ctrb}. To the best of our knowledge, our proposed survey is the only paper to review all data types and discuss more than 155 papers in the field for various types of \ac{DL} algorithms, not only \ac{CNN} and/or \ac{DNN} techniques. By presenting a thorough analysis of the proposed techniques, the paper allows researchers and practitioners to compare and contrast different approaches, identify their strengths and weaknesses, and make informed decisions on which techniques to adopt for their specific use cases. Overall, the contribution in this review paper is as follows:

\begin{itemize}
    \item Discuss the most used metrics and data sets for all data type steganalysis.
    \item Provide a strong taxonomy for  \ac{DL} algorithms used in steganalysis.
    \item Present a comprehensive coverage of existing classical steganalysis methods.
    \item Provide a comprehensive overview of the current \ac{SOTA} and offer insights into potential avenues for future research and development. Not to mention, the research gaps, challenges and open issues.
    
\end{itemize}

\begin{table}[]
\caption{Comparison of the planned study's contribution to prior steganalysis surveys. The cross mark (\xmark) denotes missing the opportunity to address the particular fields, while the tick mark (\cmark) shows that the specific field has been treated.}
\label{tab:ctrb}
\resizebox{\textwidth}{!}{ 
\begin{tabular}{lllm{0.6cm}m{0.6cm}m{0.6cm}m{0.6cm}lllm{3cm}}
\hline
\multirow{2}{*}{Reference}&\multirow{2}{*}{Type} & \multirow{2}{*}{Year} & \multicolumn{4}{c}{Media content} & \multirow{2}{*}{\begin{tabular}[c]{@{}l@{}}Coverer DL-based   \\ steganalysis article\end{tabular}} & \multirow{2}{*}{\begin{tabular}[c]{@{}l@{}}Eval. \\ metrics\end{tabular}} & \multirow{2}{*}{\begin{tabular}[c]{@{}l@{}}Databases \\ review\end{tabular}} & \multirow{2}{*}{\begin{tabular}[c]{@{}l@{}}Research gaps / Chall- \\enge / Open issues\end{tabular}}   \\
\cline{4-7}
 &  &  & Image  & Video & Audio& Text  &  &  &  \\
 \hline
Muralidharan et al. \cite{muralidharan2022infinite}&Review & 2022 & \cmark & \xmark & \xmark  & \xmark& 17 (CNN only) & \xmark & \cmark  & \cmark \\
Karampidis et al. \cite{karampidis2018review}&Review & 2018 & \cmark & \xmark & \xmark  & \xmark & 4 & \xmark & \cmark  & \xmark \\
Reinel et al. \cite{reinel2019deep}&\begin{tabular}[c]{@{}l@{}}Sys.\\  
review\end{tabular} & 2019 & \cmark & \xmark & \xmark  & \xmark & 14   (CNN only) & \xmark & \xmark & \xmark \\
Chaumont et al. \cite{chaumont2020deep}&Review & 2019 & \cmark & \xmark & \xmark  & \xmark & 24 (CNN only) & \xmark & \xmark &  \xmark \\
Hussain et al. \cite{hussain2020survey}&Survey & 2020 & \cmark & \xmark & \xmark  & \xmark & 24 (CNN only) & \xmark & \xmark & \cmark  \\
Ruan et al. \cite{ruan2020deep}&Survey & 2019 & \cmark & \xmark & \xmark  & \xmark & 15 (DNN \& CNN) & \xmark & \xmark & \cmark \\
Eid et al. \cite{eid2022digital}&Survey & 2022 & \cmark & \xmark & \xmark  & \xmark & 20 & \xmark & \cmark  & \cmark \\
Selvaraj et al. \cite{selvaraj2021digital}&Survey & 2020 & \cmark & \xmark & \xmark  & \xmark & 30 & \xmark & \xmark & \xmark \\
Bao et al. \cite{bao2021survey}&Survey & 2021 & \cmark & \xmark & \xmark  & \xmark & 12 & \xmark & \xmark & \xmark \\
This paper &Review & 2023 & \cmark & \cmark & \cmark  & \cmark & 155  & \cmark & \cmark  & \cmark \\
\hline
\end{tabular}}
\end{table}

\section{ Methodology}
\label{sec2}

\subsection{Objectives and Research Questions}
This review aims to evaluate current \ac{DL} techniques for steganalysis across diverse data types, assessing their effectiveness, resilience to adversarial attacks, and interpretability. It further explores the application of transfer learning,  reinforcement learning, assesses data and hardware limitations, and investigates privacy concerns. The review also scrutinizes the role of \ac{DL} in multi-modal steganalysis and real-time steganalysis, while identifying key areas for future research.
Table \ref{RQs} portrays the principal identified research questions
to achieve the objectives outlined above.

\begin{table*}[t!]
\caption{Research Questions.}
\label{RQs}
\small

\begin{tabular}{
m{8mm}
m{75mm}
m{75mm}
}
\hline

RQ\# & Question	& Objective   \\

\hline

RQ1 & What is the primary motivation for employing \ac{DL} in steganalysis, and what advantages it offers over traditional ML models?	& Understand the rationale behind using \ac{DL} in steganalysis and discern the unique benefits it brings as compared to traditional \ac{ML} models.  \\ \hline

RQ2 & What are the current \ac{DL} techniques used in steganalysis for different types of data? How effective are these techniques in detecting hidden messages?	& Identify and understand the state-of-the-art \ac{DL} techniques used in steganalysis for different types of data.  \\ \hline

RQ3 & How do \ac{DL}-based steganalysis methods for different types of data compare to each other and to traditional methods?	& Compare and contrast the performance of different \ac{DL}-based steganalysis methods, and evaluate them against traditional techniques.   \\ \hline

RQ4 & How can we improve the interpretability of \ac{DL}-based steganalysis models? What are the challenges and current research trends in making these models more explainable? 	& Assess the transparency of \ac{DL}-based steganalysis models, highlight the challenges in interpretability, and review the current research efforts in enhancing their explainability.   \\ \hline

RQ5 & What is the impact of adversarial attacks on \ac{DL}-based steganalysis models? How are researchers addressing the vulnerability of these models to adversarial attacks? & Investigate the susceptibility of \ac{DL}-based steganalysis models to adversarial attacks and review strategies aimed at increasing their robustness.   \\ \hline

RQ6 & How is transfer learning being applied in steganalysis, and what benefits does it offer in improving the detection of steganography across different types of data? & Explore the implementation of transfer learning in steganalysis, and evaluate its effectiveness in improving steganographic detection across different types of data.   \\ \hline

RQ7 & What are the implications of the quantity and quality of training data on the effectiveness of \ac{DL}-based steganalysis models? How can these issues be addressed?	& Explore the impact of the quality and quantity of training data on the performance of \ac{DL}-based steganalysis models, and discuss potential solutions.   \\ \hline

RQ8 & How do \ac{DL}-based steganalysis techniques affect privacy? How can these techniques be modified to respect privacy while maintaining their effectiveness? & Investigate the potential privacy concerns associated with \ac{DL}-based steganalysis techniques, and discuss possible modifications to these techniques to safeguard privacy.  \\ \hline

RQ9 & What are the key applications of steganalysis across different types of media? &  Understand the different practical uses or implementations of steganalysis across diverse media forms.   \\

RQ10 & What are the key areas that require further research in \ac{DL}-based steganalysis? What might be some potential future developments in this field? &  Identify potential future developments and key areas requiring further research in the field of \ac{DL}-based steganalysis.  \\ 

\hline

\end{tabular}
\end{table*}

\subsection{Quality assessment}
Our study conducted a quality assessment using a questionnaire, which included eight questions relevant to the research questions and objectives. These questions covered topics such as (i) the clarity and appropriateness of the research questions, (ii) the comprehensiveness of the literature review, (iii) the appropriateness and relevance of the datasets used, (iv) the effectiveness of the classification methods used, (v) the depth and comprehensiveness of the discussions on proposed solutions, (vi) the degree to which experimental analysis supports the proposed solutions, and (vii) the clarity of the presentation of the results. By answering these questions, we evaluated the quality and identified the advantages and disadvantages of each work. The formulation of the quality assessment criteria in Table \ref{tab:quality} was based on eight questions that assessed the quality of our research in relation to the research questions.

\begin{table*}[t!]
\caption{Research Questions.}
\label{tab:quality}
\small

\begin{tabular}{
m{5mm}
m{65mm}
m{65mm}
m{25mm}
}
\hline

No. & Question	& Objective  & Mapping to the research question\\

\hline

Q1 & Were the research questions appropriately presented and well-articulated?	& 
The clarity and appropriateness of the research questions should be evaluated. & RQ1, RQ2 \\ \hline

Q2 & Was a comprehensive review of the state-of-the-art in the relevant areas provided by the authors? & The comprehensiveness of the literature review should be evaluated. & RQ2, RQ3, RQ4 \\ \hline

Q3 & Was the appropriateness and effectiveness of the methods used for classifying the studies, datasets, models, and attacks confirmed?	& The appropriateness and effectiveness of the classification methods and taxonomy used should be reviewed. & RQ4, RQ5, RQ6 \\ \hline

Q4 & Was the appropriateness and relevance of the datasets discussed in the study confirmed?	& The suitability and significance of the datasets used in the study should be evaluated. & RQ7 \\ \hline

Q5 & Has a detailed discussion on the potential benefits, challenges, and implications of the overviewed solutions been carried out?	& Assess the depth and comprehensiveness of the overviewed solutions  & RQ1, RQ2, RQ3, RQ4, RQ6 \\ \hline

Q6 & 
Has a discussion been carried out on the limitations of the overviewed solutions?	& Assess the constraints of the overviewed solutions. & RQ7, RQ8 \\ \hline

Q7 & Were the potential applications of the overviewed solutions discussed in detail?	& 
The thoroughness of the discussions on the potential applications of the proposed solutions should be reviewed. & RQ9 \\ \hline

Q8 & Were the study's conclusions and recommendations derived from the presented results and analysis?	& Assess the extent to which the results and analysis substantiate the study's conclusions and recommendations. & RQ9, RQ10 \\ \hline

\end{tabular}
\end{table*}

\subsection{Study selection criteria }

In order to create a comprehensive list of references for steganalysis based on DL, several well-known journals databases such as Scopus and Web of Science, including IEEE Xplore, ACM Digital Library, Science Direct, and SpringerLink, were searched. The selection criteria for these publications were based on four factors: (i) \textbf{modularity,} which kept only the most common methods and implementations; (ii) \textbf{coverage,} which is reported on a new or specific application domain; (iii) \textbf{influence,} which limited the selection to works published in high-quality journals, book chapters, or conference proceedings, but with a high number of citations; (iv) and \textbf{newness}, which only included papers published recently. 

\begin{figure}[!t]
    \centering
    \includegraphics[scale=0.4]{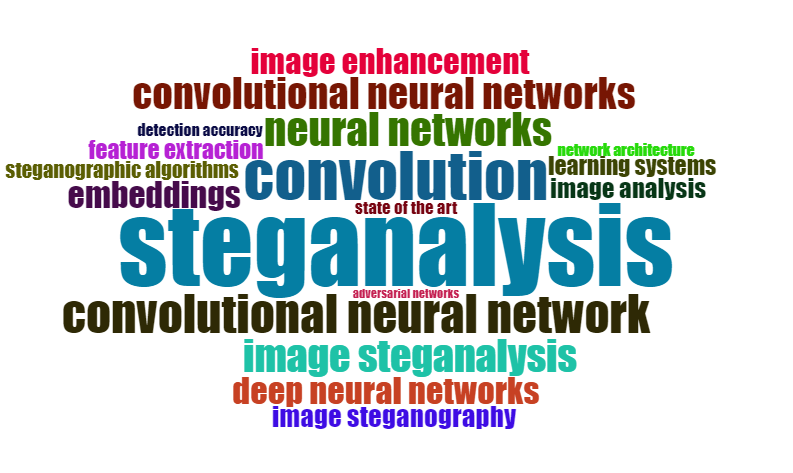}
        \caption{ Most frequently keywords used by steganalysis researchers.}
    \label{keywords}
\end{figure}

\ac{DL}-related keywords were used to conduct the initial search for steganalysis. The papers were then divided into five groups based on "topic clustering" using the keywords associated with the application field. Figure \ref{keywords}, which has been generated using the biblioshiny  analysis tool\footnote{\url{https://www.bibliometrix.org/home/index.php/layout/biblioshiny}}, illustrates the most frequent keywords used by steganalysist in their research papers. We can easily distinguish that image steganalysis is the most conducted field compared to other types of carriers. Also, \ac{CNN} is the most frequently used technique in steganalysis.

\begin{figure}[!t]
    \centering
    \includegraphics[scale=0.6]{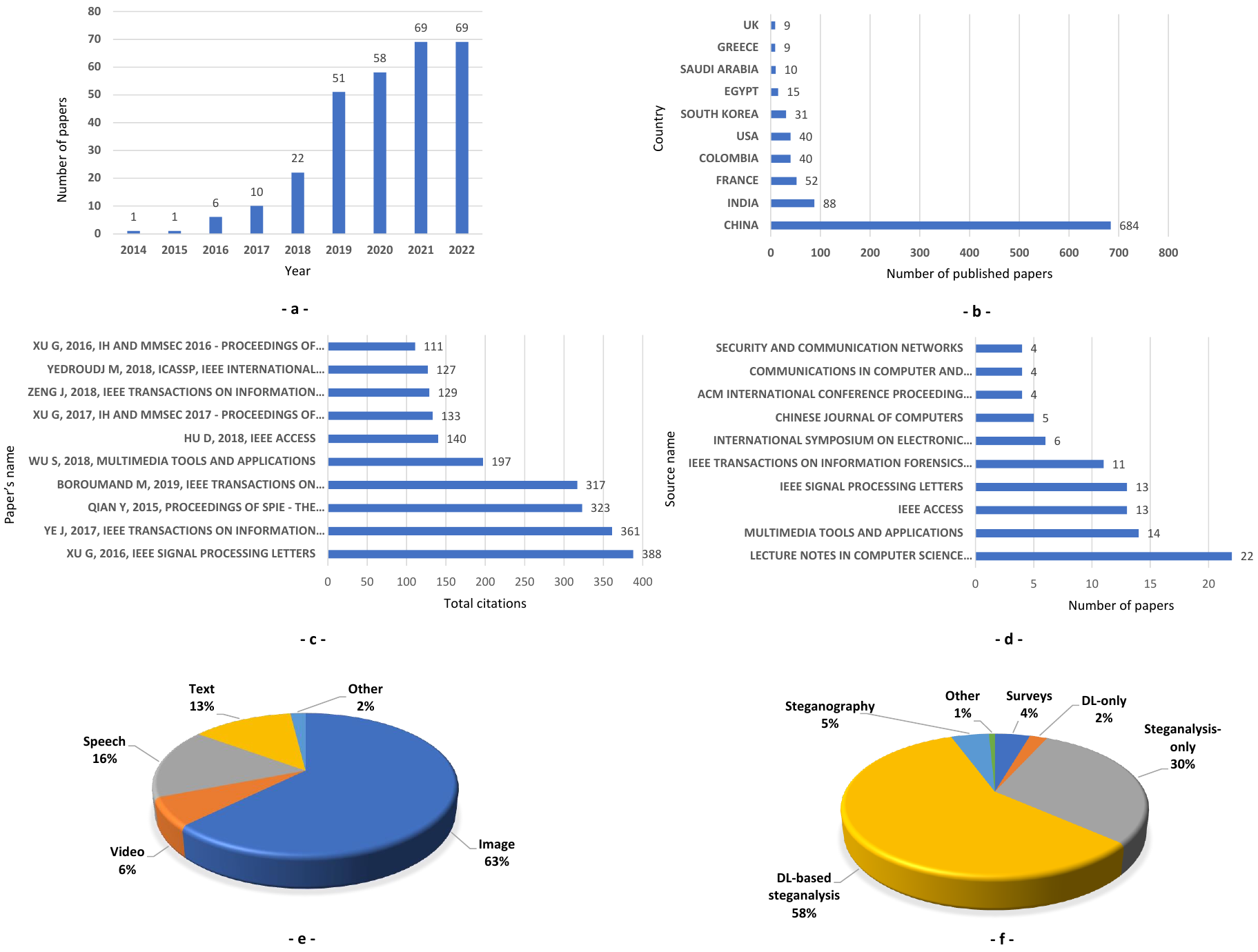}
    \caption{ Bibliography analysis. (a) Annual production, (b) Top 10 production country, (c) Top 10 most global cited documents, (d) Top 10 most relevant sources, (e) Comparison based on data-type DL steganalysis, (f) Comparison based on research paper's type used in this review. }
    \label{statistics}
\end{figure}

\subsection{Bibliometric statistics}
Figure \ref{statistics} displays the statistics obtained after applying the selection criteria, including the number of articles involved in this review each year, the top ten most cited papers with their authors and publication year, the top ten most relevant papers' sources, the top ten most active authors, the percentage of papers for DL-based steganalysis, steganalysis only, DL only, surveys, and others, and the percentage of discussed papers in each field of DL-based steganalysis.

\section{Background in \ac{DL} and  steganalysis}
\label{sec3}
\subsection{Metrics}

The metrics that are frequently used to evaluate steganalytic methods, and the functionalities they assess, are explained below (Table \ref{tab:metrics}). In this table, \Ac{TP} is the number of carriers that are correctly classified as stego (i.e., containing an embedded message) and \Ac{FN} is the number of carriers that are identified as cover (i.e., not embedded), but they actually contain a hidden message. Conversely, \Ac{TN} and \Ac{FP} are, respectively, unembedded carriers that are identified as cover (correctly) and stego (incorrectly). By using these metrics, researchers can objectively evaluate the effectiveness and limitations of different steganalysis methods and identify areas for improvement.

\begin{table}[]
\begin{tabular}[!t]{m{40mm}m{40mm}m{90mm}}
\small
\caption{A summary of the evaluation metrics used to assess steganalysis techniques.}
\label{tab:metrics} \\
\hline
Metric & Formula & Description   \\ 
\hline
\Ac{FPR} and \Ac{TPR} & \(\displaystyle \mathrm{\frac{FP}{FP+TN}}, \mathrm{\frac{TP}{TP+FN} }\) &  The \Ac{FPR}, also known as \ac{FAR}, is the ratio (or percentage)  of the non-steganographic bearers that are incorrectly identified as containing hidden information.  The \Ac{TPR}, also known as \ac{DR}. is the ratio (or percentage) of the steganographic bearers that is correctly identified as containing hidden information. \\ [0.6cm]  


\Ac{MSE} &\(\displaystyle
\frac{\sum_{M,N}[I_1(m,n)-I_2(m,n)]^2}{M\cdot N} \)& The \Ac{MSE} is the cumulative squared error between the stego image and the original cover image. $M$ and $N$ are the rows and columns,  respectively. $I_1(m,n)$ and $I_2(m,n)$ are the values of the pixel in the position $(m,n)$ for the cover and the stego images. \Ac{MAE}=$\frac{1}{N}\sum_{i=1}^N|x_i-y_i|$ used to gauge the precision of estimating the embedding rate.\\[0.6cm]
\Ac{PSNR}& \(\displaystyle
\mathrm{10 \log_{10}\Big(\frac{\mathit{R}^2}{MSE} \Big)} \) &  $R$ is the fluctuation in the input image, i.e., the maximum difference for the values of the pixels. The \ac{PSNR} is commonly used to evaluate the quality of the stego-image (the image after the hidden information has been embedded). A lower \ac{PSNR} indicates that the steganography technique has introduced more distortion to the image.\\ [0.6cm]
\Ac{SSIM}& \(\displaystyle \mathrm{\frac{(2\mu_x\mu_y+c_1)(2\sigma_{xy}+c_2)}{(\mu_x^2+\mu_y^2+c_1)(\sigma_x^2+\sigma_y^2+c_2)}} \) &  The \Ac{SSIM} is a measure of the similarity between two images. A lower \ac{SSIM} indicates that the steganography technique has introduced more distortion to the image. The details of the parameters of the formulae can be found in \cite{wang2004image}.\\ [0.6cm]



Precision and recall& \(\displaystyle \mathrm{\frac{TP}{TP+FP}}  \), \(\displaystyle \mathrm{\frac{TP}{TP+FN}} \) &  Precision is the ratio of correctly identified carriers of hidden information among the total number of instances identified as stego by the steganalysis algorithm. Recall is the ratio of correctly identified carriers of hidden information among the total number of hidden information instances present in the data. High precision and recall values indicate that the steganalysis algorithm effectively detects hidden information in the data.\\ [0.8cm]


\Ac{DER}&\(\displaystyle \frac{1}{2} (P_{\mathrm{FA}}+P_{\mathrm{MD}})  \)& $P_{\mathrm{FA}}$ and $P_{\mathrm{MD}}$ are the false-alarm and missed-detection probabilities, respectively. The lower the \ac{DER}, the more effective the steganalysis method. Note that $\ac{DER}$ can be expressed as $1 - \mathrm{Accuracy}$.\\
 \hline
 \end{tabular}
\end{table}

\subsection{Data sets}

Steganalysis data set refers to a collection of data used for training and evaluating models that are designed to detect the presence of hidden messages or information within digital media, such as images, audio, or videos.

Table \ref{tab:datasets} presents a list of commonly used steganalysis datasets, all of which have been exclusively used in the context of deep learning-based steganalysis in at least one work or research article. It is important to note that these datasets have gained varying levels of popularity within the steganalysis field classified in table as "High," "Moderate," "Low," and "Minimal." In most cases, the datasets are  primarily contain cover samples only, as indicated in the table by a cover/stego ratio of 1:0.  Instead, Stego samples, which are essential for steganalysis, are generated using steganographic methods like HUGO, WOW, UNIWARD, and others applied to the cover samples. It is worth pointing out that the steganographic methods used in the works are described in columns "Steganographic algorithm" of Tables  \ref{TCoverBased}, \ref{tab:speechText}, and \ref{tab:techType}. Moreover, the labeling of these datasets
encompasses both cover and stego classes, or multiple stego classes in cases where several steganographic algorithms are used to generate stego samples. Additionally, it is noteworthy that some datasets lack information about the division between training and test samples, as indicated by ’–’ in the table.

\begin{table*}[ht]
\caption{Analyzing current Internet datasets for the evaluation of DL-based steganalysis capabilities.}
\label{tab:datasets}
\begin{tabular}{m{2cm}m{1cm}m{1cm}m{1.5cm}m{1.5cm}m{1.5cm}m{1cm}m{1cm}m{1cm}m{3cm}}
\hline
 Data set & Year & Carrier \newline type & Popularity & \# Samples &   \# Training \newline samples& \# Test \newline samples& Stego/ \newline Cover ratio & Data \newline
labelled?  &  \ac{DL}-based works
using the data set  \\
 \hline
 
 BOSS& 2011  & Image & High &  11,000 &10,000 (BOSSBase)&1000 (BOSSRank)& 1:0  &  \xmark & \cite{kato2020preprocessing, ye2017deep, lu2019importance, qian2015deep, pibre2015deep, xu2016structural, xu2017deep, qian2018feature, yang2018analysis, mustafa2020enhancing, jung2021pixelsteganalysis, zeng2019wisernet, lai2022generative, you2019restegnet, huang2022image, liu2023image, boroumand2018deep, you2020siamese, arivazhagan2021hybrid, yousfi2020intriguing, chen2017jpeg, zhang2019deep, wang2021steganalysis, kang2021identification, zhang2022generative, weng2022lightweight, yang2017jpeg, yang2018deep, prasad2020detection, ghosh2022deep, plachta2022detection, zhang2020cover, deng2022universal, zhang2019new, deng2021spatial, zhang2022dataset, zhan2017image, dengpan2019faster, el2018improved, mustafa2019accuracy, yang2020reinforcement, ni2019selective,  cohen2020assaf, kang2020classification} \\
 
 BOWS2 & 2008 & Image & High &  10,000 &-- &--& 1:0  &  \xmark &\cite{ye2017deep,huang2022image, yousfi2020intriguing, chen2017jpeg, weng2022lightweight, yang2017jpeg, deng2022universal, deng2021spatial, yang2020reinforcement, hu2019digital} \\
 
 IStego100K & 2020  & Image & Low & 208,104 &200,000&8,104& 1:1  &  \cmark &\cite{lai2022generative, ahmed2022image} \\
 ALASKA2 & 2020  & Image & Moderate &   305,000 &300,000&5000& 1:4  &  \cmark  &\cite{you2020siamese, padmasiri2021impact, yousfi2020imagenet} \\
 
ImageNet & 2009  & Image & Moderate & 14,197,122 & &&1:0  &  \xmark  &\cite{mustafa2020enhancing, jung2021pixelsteganalysis, zeng2017large, yang2017jpeg, qin2022feature, yang2018deep, bonnet2021forensics, zhang2022adnet}\\

Cifar-10 & 2009  & Image & Low & 60,000 & 50,000 &10,000&1:0  & \xmark  & \cite{jung2021pixelsteganalysis, qin2022feature} \\
COCO &  2015 & Image & Low & 330,000 & -- &--& 1:0  & \xmark &    \cite{zhang2019enhancing, liu2022feature} \\

LIRMMBase &  2015 & Image & Minimal & 16,580 & -- &--& 1:0  & \xmark &    \cite{zhang2019deep} \\

Raise & 2015  & Image & Minimal & 8,156 & -- &--& 1:0  & \xmark &    \cite{prasad2020detection} \\

LFW & 2007  & Image & Minimal & 13,233 & -- &--& 1:0  & \xmark &    \cite{ghosh2022deep} \\

UCID & 2003  & Image & Minimal & 1338 & -- &--& 1:0  & \xmark &    \cite{zhang2022dataset} \\

MIRFlickr25K & 2008  & Image & Minimal & 25,000 & -- &--& 1:0  & \xmark &    \cite{zhang2022dataset} \\

DIV2K & 2017  & Image & Minimal & 1,000 & 800 & 100 & 1:0  & \xmark &    \cite{zhang2022dataset} \\

Xiph video test &  -- & Video & Minimal & 120 & -- &--& 1:0  & \xmark &    \cite{huang2020deep} \\
media

TIMIT& 1986  & Speech & Low &   6300 &4620 &1690 &1:0  &  \xmark &   \cite{lee2020deep, lin2019audio} \\

Noiseus & 2007  & Speech & Minimal & 30 & & &1:0  &  \xmark &   \cite{paulin2016audio} \\

 Wall Street Journal & 1992  & Speech & Minimal &   86000 &--&--& 1:0 & \xmark &\cite{chen2017audio} \\

 IMDB Movie Review & 2011 & Text & Moderate &   50,000 & 25,000&25,000& 1:0  &  \xmark  & \cite{xu2021small, li2021text, niu2019hybrid,  yi2022exploiting, fu2022hga, peng2021real} \\

News & 2022 & Text &Low & 210,294 & --&--& 1:0  &  \xmark  & \cite{xu2021small, li2021text} \\

Twitter & 2009 & Text & Moderate & 1,600,365 & 1,600,000 &365& 1:0  &  \xmark  & \cite{xu2021small, li2021text, niu2019hybrid, yi2022exploiting, fu2022hga} \\

Gutenberg & 2013 & Text & Minimal & 3,036 & --&--& 1:0  &  \xmark  & \cite{niu2019hybrid} \\

Human UCSC-hg38 & 2013 & DNA  & Low & 24,279 (genes) & --&--& 1:0  &  \xmark  & \cite{bae2018dna, bae2020dna} \\

 \hline
\end{tabular}
\end{table*}

\subsection{Taxonomy of \ac{DL}}

\Acp{DNN} can automatically obtain feature representations for steganographic detection through sample training, eliminating the need for manually-defined features. The key challenge now is to design the structure of \Acp{DNN}. Speech, text, video, and image steganalysis requires the extraction of subtle steganography information features hidden behind payload content and texture, which is quite different from traditional steganalysis tasks. Increasing the signal-to-noise ratio and maximizing the residual information are usually necessary to improve the performance of steganographic detection. For that reason, many \ac{DL} techniques play a great role in successfully detecting or even uncovering the steganogram. The most used DL-based steganalysis techniques are overviewed in the next few sections. Table \ref{tab:DLcmp} compares various \ac{DL} techniques for steganalysis in different data types. It assesses their suitability and effectiveness, with categories ranging from "Excellent" and "Good" to "Limited" and "Promising." The architectures include CNN, LSTM, DNN, RNN, DBN, RBN, and GNN, applied to text, speech, image, and video steganalysis. The table helps users understand the relative strengths and weaknesses of these architectures for specific tasks.


\begin{table}[H]
\centering
\scriptsize
\caption{Comparison of DL techniques for diverse data-type steganalysis.}
\label{tab:DLcmp}
\begin{tabular}{m{1.2cm}m{3.5cm}m{3.5cm}m{3.5cm}m{3.5cm}}
\hline
DL type &  Text Steganalysis & Speech Steganalysis & Image Steganalysis & Video Steganalysis \\
\hline
CNN & Excellent for text \newline steganalysis & Limited applications in \newline  speech & {Excellent} for image \newline  analysis & {Promising} for analyzing \newline  video \\
\hline
LSTM & Good for sequential \newline data & {Good} for analyzing \newline speech & Limited applications for \newline images & Limited application for \newline videos \\
\hline
DNN & {Good} for various \newline data types & {Good} for various \newline data types & Limited applications for \newline images & Limited application for \newline videos \\
\hline
RNN & {Good} for sequential \newline data & {Good} for sequential \newline data & Limited applications for \newline images & Limited applications for \newline videos \\
\hline
DBN & Limited applications for \newline text analysis & Limited applications for \newline speech analysis & Limited applications for \newline image analysis & Limited applications for \newline video analysis \\
\hline
RBN & Limited applications for \newline text analysis & Limited applications for \newline speech analysis & Limited applications for \newline image analysis & Limited applications for \newline video analysis \\
\hline
GNN & Limited applications for \newline text analysis & Limited applications for \newline speech analysis & Limited applications for \newline image analysis & Limited applications for \newline video analysis \\
\hline
\end{tabular}
\end{table}

\subsubsection{DNNs}
\ac{DNN} is a type of artificial neural network that has more hidden layers compared to shallow counterparts. While there may be variations in specific designs based on different problem requirements, the \ac{DNN} model typically consists of an input layer, an output layer, and multiple hidden layers. This kind of architecture is suitable for solving classification and regression problems, but the size of the model parameters increases with the number of input features, which can affect the computational performance. To learn the \ac{DNN} model, the back-propagation learning algorithm can be used to adjust the weights by propagating errors from the output layer to the preceding layers. However, tuning the hyper-parameters of the model is crucial for optimal performance. Thus, determining the best model parameters is a significant challenge \cite{catal2022applications}.

\subsubsection{CNNs}
\ac{CNN} utilizes convolution layers that rely on the convolution operation to identify hidden patterns in images. This process was first introduced to detect low-level features before gradually detecting more complex features . Lower layers in the network are responsible for identifying basic features, while higher layers identify complex features \cite{himeur2023face}. The \ac{CNN} model is most commonly used for image classification tasks. The \ac{CNN} architecture consists of four different types of layers: convolution layers, max-pooling layers, dropout layers, and \ac{MLP} layers. In \ac{CNN} architecture, the \ac{MLP} layer is fully connected. The basic convolution operation can be expressed using Equation \ref{conv2D}, where $p$ represents the input, $t$ represents time, $k$ represents the kernel,  and \(f(t, x, y)\) represents a feature map.

The variables \(u\) and \(v\) represent the spatial coordinates in the input signal, and \(t - u\) and \(x - v\) represent the relative displacement between the kernel and the input signal at each spatial position. The \(y\) coordinate is preserved throughout the convolution.


\begin{equation}
 f(t, x, y) = f(p \ast k)(t, x, y) = \sum_{u = -\infty}^{\infty} \sum_{v = -\infty}^{\infty} p(u, v) \cdot k(t - u, x - v, y)
 \label{conv2D}
\end{equation}

In this equation,  represents the output signal at time (\(t\) and spatial position \((x, y)\), \(p(u, v)\) is the input signal in spatial position \((u, v)\) and \(k (t - u, x - v, y)\) is the 3D kernel. The summation is performed over all possible values of \(u\) and \(v\), which can range from negative infinity to positive infinity. However, in practical applications, the signals are often discrete, so the summation is performed over a finite range determined by the size of the input signal and the kernel.


Equation \ref{2d_nn} shows a neural network architecture where $w$ represents weight, $I$ represents input, $b$ represents bias, and $y$ represents the output of the neuron. Once the output of the neuron is obtained, the softmax function is applied to obtain the final output. Equation \ref{2d_softmax} given in \cite{zhu2020efficient}.

\begin{equation}
 y_{i,x,y} = \sum_{j=1}^{N} w_{i,j} \cdot I_{j,x,y} + b_{i},
 \label{2d_nn}
\end{equation}

\begin{equation}
\mathrm{softmax}(y_{i,x,y}) = \frac{e^{y_{i,x,y}}}{\sum_{j} e^{y_{j,x,y}}}.
\label{2d_softmax}
\end{equation}

The softmax function computes the exponential of the input value \(y_{i,x,y}\) and normalizes it by the sum of the exponentials of all input values across the neurons (\(j\)) at the same spatial position \((x, y)\). This normalization ensures that the resulting values lie in the range of [0, 1] and sum up to 1, making them suitable for representing probabilities.




\subsubsection{RNNs}

\begin{figure}[h!]
    \centering
    \includegraphics[scale=0.45]{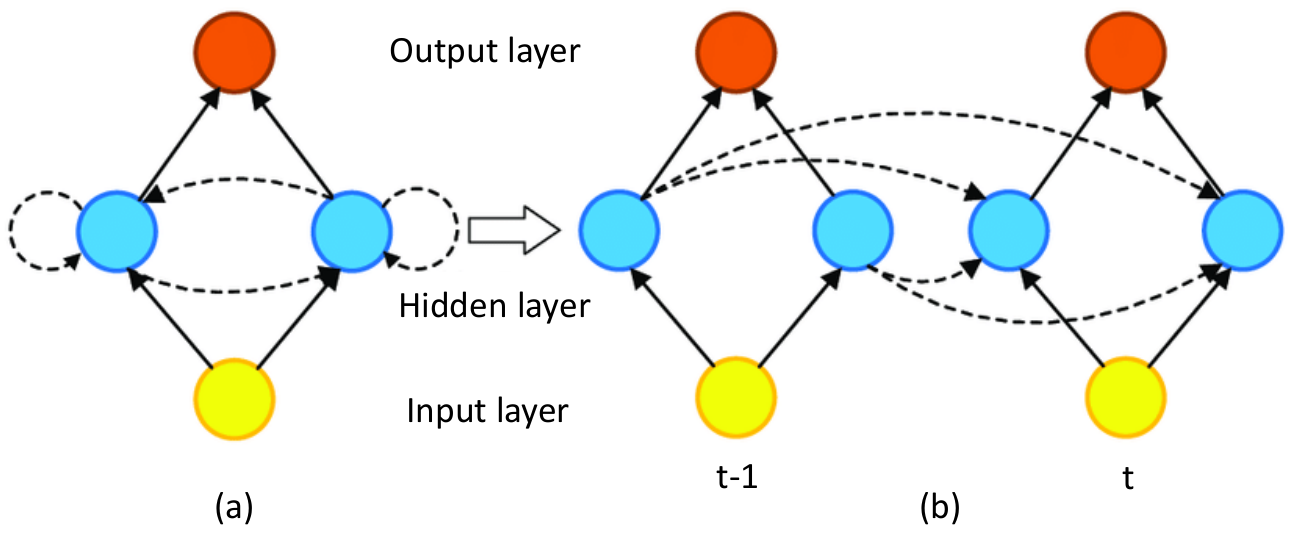}
    \caption{ The architecture of: (a) RNN and (b) RNN across a time step.}
    \label{RNN}
\end{figure}

\ac{RNN} is a type of \ac{NN} that uses the output from the previous stage as input in the current stage [19]. Unlike conventional neural networks, where inputs and outputs are independent, \acp{RNN} are useful in scenarios where past inputs are needed to predict future outputs, such as predicting the next word in a sentence. \acp{RNN} use a hidden layer to remember past inputs, making it different from the \Ac{DNN} model in terms of input processing, as shown in Figure \ref{RNN}. 

The hidden state is the essential component of the \ac{RNN} model, which retains information about past input sequences. The \ac{RNN} model comprises several layers and units, with the hidden layer unit containing information about the input history in the ``state vector''. The \ac{BPTT} algorithm can be used to learn the \ac{RNN} model, and the \ac{SGD} and \ac{RMSProp} algorithms can be used to optimize \ac{RNN} parameters. However, \ac{RNN} training is more challenging than other models due to its reliance on time, and its complexity increases with the learning period. The primary purpose of utilizing an \ac{RNN} model is to learn long-term dependencies, although it has been found in the literature that learning \acp{RNN} can be challenging for long-term dependencies.

\subsubsection{LSTMs}
\ac{LSTM} is an enhanced version of the \ac{RNN} model that addresses the issue of \acp{RNN} failing in certain circumstances \cite{hochreiter1997long}. Unlike the RNN model, where the output of the previous stage is used as input in the current stage, the \ac{LSTM} network is capable of storing past information for a longer period, making it suitable for applications requiring long-term dependencies \cite{wu2016google}. The \ac{LSTM} network consists of \ac{LSTM} units that are combined to form an \ac{LSTM} layer. Each \ac{LSTM} unit is composed of cells that have input, output, and forget gates that control information flow within the network, allowing each cell to remember important information for an extended period. Equations \ref{lstm} represents the forward pass of an \ac{LSTM} unit \cite{catal2022applications}.

\begin{equation}
\begin{split}
A_f=S_f\Big(W_f*L_i+U_f*L_{j-1}+b_f\Big), \hspace{0.7cm} A_i=S_f\Big(W_i*L_i + U_i * L_{j-1} + b_i\Big), \hspace{0.7cm} 
 L_j = A_j * H_f(V_c)  \\
A_j = S_f\Big(W_0*L_i+ U_0* L_{j-1}+ b_j\Big), \hspace{1cm} 
 V_c = A_f*V_{c-1}+ A_i * C_f \Big(W_c
* L_i + U_c * L_{j-1} + b_c\Big) \\
\end{split}
\label{lstm}
\end{equation}


In this context, the symbols $L_i$ and $L_j$ stand for the input and output of the \ac{LSTM} unit, respectively. $A_f$, $A_i$, and  $A_j$  denote the activation vectors for the forget, input, and output gates, respectively. Additionally, $V_c$ is used to indicate the cell state vector, while $S_f$ represents a \textit{sigmoid} function. The hyperbolic tangent function is represented by $H_f$ and $C_f$. The weight matrices are denoted by $W$ and $U$, while the bias vector is represented by $b$ \cite{greff2016lstm}.

\subsubsection{Auto encoders}

\begin{figure}[h!]
    \centering
    \includegraphics[scale=0.9]{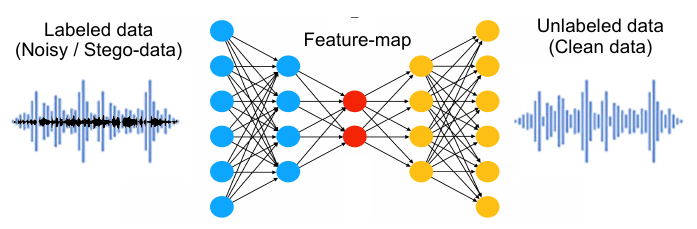}
    \caption{ The general layout of a \ac{AE}.}
    \label{dae}
\end{figure}

\Acp{AE} is a type of neural network that consists of two components, an encoder, and a decoder. The encoder takes in input data, such as an image, and compresses it into a compact latent representation. This compression process helps to identify the important features of the input (also known as feature-map). The decoder then takes the latent representation and converts it back into the original data in an unsupervised manner (unlabeled target). Essentially, the auto-encoder performs a compression-decompression operation, it is also useful for denoising, and steganogram removal. Figure \ref{dae} illustrates a general architecture of an auto-encoder \cite{kheddar2022high}.

\subsubsection{RBMs}
\begin{figure}[h!]
    \centering
    \includegraphics[scale=0.85]{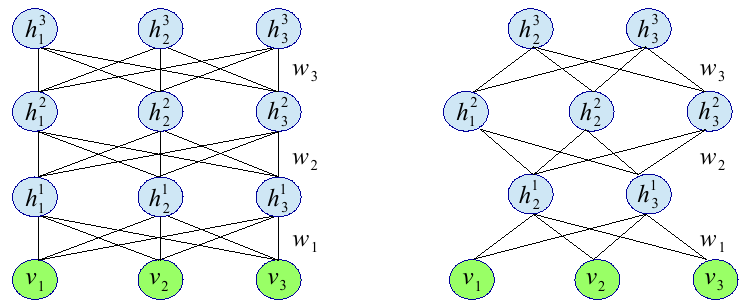}
    \caption{ Architecture of: (a) \ac{RBM} and (b) \ac{DBN} networks.}
    \label{rbmRbn}
\end{figure}

\acp{RBM} are a category of unsupervised learning models used
to  determine the probability distribution of input data \cite{qiu2014ensemble}. They are capable of discovering hidden patterns without supervision, but the training process can be challenging, which is a drawback. \acp{RBM} are composed of visible (denoted $v_i$) and hidden layers (denoted $h_i^j$) that form a bipartite and undirected graph, with no connections between layers. Each cell in the network processes input data and decides whether to transmit it or not, after multiplying it with weights and adding bias. The activation function is applied to the resulting input, generating an output that re-enters the network in the reconstruction phase. In Figure \ref{rbmRbn} (a), the neurons in both the visible and hidden layers form a bipartite graph, where each visible neuron is connected to all the hidden neurons, but the neurons within the same layer are not connected. The Gibbs sampler technique is used to train an \ac{RBM} model to minimize the log-likelihood of both the data and the model \cite{catal2022applications}.

\subsubsection{DBNs}

A \ac{DBN} is a generative artificial neural network that build upon the principles of RBMs. This kind of network consists of multiple layers of interconnected, undirected RBMs. These layers are usually divided into two primary types: the visible layer denoted as $v_i$ and the hidden layers represented as $h_i^j$ (as illustrated in Figure \ref{rbmRbn} (b)). \acp{DBN} are trained in an unsupervised manner using an algorithm called ``layer-wise pre-training''. The output of each layer is used as input for the next layer, creating a deep network that can learn high-level representations of the input data. To train the \ac{DBN} model, there are two steps involved. The first step is stacked \ac{RBM} learning, which uses the iterative \ac{CD} algorithm, and the second step is back-propagation learning, which applies different optimization algorithms. In other words, the last layer of the \ac{DBN} is usually trained using a supervised learning algorithm, such as back-propagation, to fine-tune the network for a specific task.  \acp{DBN} have been used for various applications including image classification, speech recognition, and anomaly detection. Both the \ac{RBM} and \ac{DBN} models have the same hyper-parameters, which can be optimized using the \ac{CD} algorithm.
Like the \ac{RBM} model, the \ac{DBN} model has a disadvantage in that it first learns the input data in a probabilistic manner and then identifies independent features. After unsupervised learning, the classification is performed through supervised learning.


\subsubsection{GNN}

\ac{GNN} is a type of neural network which directly operates on the graph structure. A typical application of \ac{GNN} is node classification. Essentially, every node in the graph is associated with a label, and we want to predict the label of the nodes without ground truth. This section illustrates the algorithm described in the paper, the first proposal of \ac{GNN} and thus is often regarded as the original \ac{GNN}. In the node classification problem setup, each node $v$ is characterized by its feature $x_v$ and associated with a ground-truth label $t_v$. Given a partially labeled graph $G$, the goal is to leverage these labeled nodes to predict the labels of the unlabeled ones. This architecture learns to represent each node with a $d$ dimensional vector (state) $h_v$ which contains its neighborhood information. Specifically,

\begin{equation}
    h_v=f\left(x_v,x_{co[v]},h_{ne[v]},x_{ne[v]}\right)
\end{equation}

\noindent Where $x_{co[v]}$ denotes the features of the edges connecting with $v$, $h_{ne[v]}$ denotes the embedding of the neighbouring nodes of $v$, and $x_{ne[v]}$ denotes the features of the neighbouring nodes of $v$. The function $f$ is the transition function that projects these inputs onto a $d$-dimensional space \cite{scarselli2008graph}.

\subsection{Taxonomy of steganalysis methods}

Steganalysis methods can be categorized into two types: specific (or targeted) and universal steganalysis. Targeted steganalysis is effective in detecting specific steganography algorithms, with a low false alarm rate and accurate results, but its practical application is limited. Examples of specific steganalysis algorithms include regular-singular analysis \cite{lou2009message}, raw quick pair analysis \cite{fridrich2002practical}, and blockiness artifacts analysis \cite{baziyad2020achieving}. Universal steganalysis treats steganographic detection as a classification problem and uses \ac{ML} to extract high-dimensional features for classification. Examples of universal steganalysis methods include subtractive pixel adjacency matrix feature analysis \cite{sahu2021multi}, steganalysis of JPEG images based on Markov features \cite{he2012digital}, and spatial rich model feature extraction \cite{wang2016pure}. Although these methods significantly improve detection performance, they require longer training times due to the use of high-dimensional features. Feature design is a critical element in steganalysis, and the features used in the model are often obtained through manual design. This requires professional knowledge and substantial manual intervention, and the quality of the manually defined features directly affects the performance of the model.

\begin{figure}[h!]
    \centering
    \includegraphics[scale=0.6]{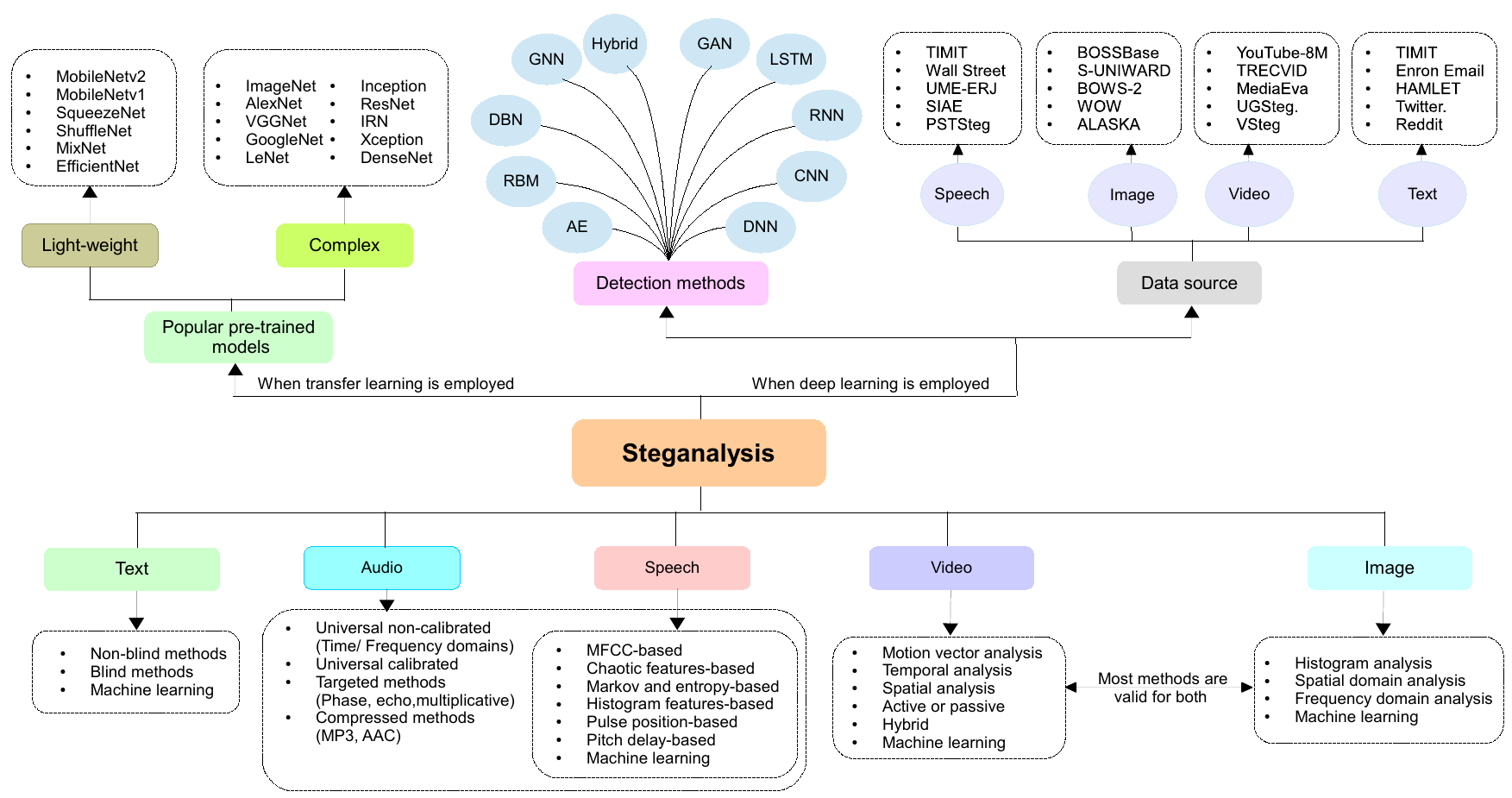}
    \caption{ Taxonomy of existing steganalysis methods.}
    \label{TaxonomySteg}
\end{figure}

There are different techniques for steganalysis, depending on the type of data being analyzed. A brief taxonomy of \ac{DL}-based steganalysis techniques for image, video, speech, and text data is presented in Figure \ref{TaxonomySteg}.

A brief overview of the different techniques depending on the type of the carrier is provided below:
\begin{itemize} 
\item \textbf{Image and video steganalysis:} 
There are several techniques for image steganalysis, including histogram analysis \cite{lu2019binary,abdali2020reference}, spatial domain analysis \cite{ren2020learning,chubachi2020ensemble,wang2020towards}, and frequency domain analysis  \cite{khalifa2019image}. 
The researchers approach the video as a collection of images without taking into account the unique temporal characteristics of the video domain, thus many image steganalysis techniques are valid for video steganalysis. According to the review  \cite{bouzegza2022comprehensive}, video steganalysis techniques can be specific or targeted steganalysis, also known as passive steganalysis, which rely on prior knowledge of the steganography tool or embedding algorithm employed to conceal the confidential message  \cite{shi2021hevc}. Active steganalysis is the process of detecting or extracting a concealed message with limited or no prior information  \cite{ghamsarian2021blind}.  Video steganalysis techniques can be also temporal  \cite{weng2019high,wu2020sstnet}, spatial like the methods surveyed in \cite{dalal2018video}, or hybrid methods  \cite{tasdemir2015spatio,wu2020sstnet}. Video steganalysis can rely on motion vector analysis, this method examines the motion vectors in the video to detect any changes or anomalies that may indicate the presence of hidden data  \cite{zhai2019universal,huang2020combined}.
     
\item \textbf{Speech steganalysis:} 
As per the findings of \cite{ghasemzadeh2018comprehensive}, in the context of speech/audio steganalysis, there are non-targeted and non-compressed techniques that are classified as either spectral-like  method-based \ac{FFT} \cite{peng2019fast}, method based \ac{MFCC} \cite{ghasemzadeh2016audio} or temporal methods like \ac{LSB} 
 method \cite{yang2017steganalysis}, pitch delay \cite{ren2016amr}, pulse positions \cite{ren2015amr}. Alternatively, calibrated methods include denoising-based \cite{ozer2006detection}, constant referencing \cite{yazdanpanah2022monitoring}, and cover linear basis or re-sampling techniques \cite{hassaballah2021novel}. For targeted approaches, speech/audio steganalysis can be based on phase, echo, or multiplications \cite{yazdanpanah2022monitoring}. Additionally, speech/audio steganalysis can also be applied to compressed modes \cite{hassaballah2021novel,miao2014steganalysis}. 

\item \textbf{Text steganalysis:} Discovering secret information that is concealed within text documents is already summarized in \cite{yang2022overview} and it includes non-blind methods such as distribution of first letters of words analysis \cite{sui2006steganalysis}, context information analysis \cite{wang2013novel}, analysis-based evolutionary algorithm \cite{din2013fitness}, synonym frequency analysis \cite{xiang2014linguistic}.  Blind text steganalysis methods include techniques based on statistical language models  \cite{guo2022linguistic,meng2009linguistic}.
 
\end{itemize}

\Ac{ML} can be used as steganalysis method that involves training a model using a large data set of images, videos, speeches, or text documents, with the aim of recognizing patterns of hidden data in new images, videos, speeches, or text files. This approach is especially useful in steganalysis, as it enables the model to identify subtle changes or anomalies in the data that may indicate the presence of concealed information  \cite{shankar2023random}.

While various deep learning models, such as DNNs, CNNs, and LSTMs, are commonly applied in steganalysis across these diverse data types, it is essential to emphasize that, despite the shared core deep learning models, there exist significant distinctions in the methodologies and considerations associated with each data modality:

\begin{itemize}
\item \textbf{Data Preprocessing:} Each data modality requires unique preprocessing techniques tailored to its inherent characteristics. For instance, image steganalysis involves pixel-level analysis, whereas speech steganalysis necessitates audio feature extraction methods like MFCCs. These distinct preprocessing steps contribute to the diversity in steganalysis methodologies.
\item \textbf{Network Architectures:} Generally, DNNs are versatile models used for various tasks in data analysis, such as classification. CNNs are primarily employed to process 1D data for tasks related to text and speech. They can handle also 2D and 3D data for images and videos. RNNs and LSTMs excel in retaining information from previous states, which is valuable for analyzing sequential data such as text and speech. Furthermore, GANs outperform in generating duplicate data, contributing to acquiring more information and addressing data scarcity issues. Similar to GANs, RBMs serve as fundamental components in constructing complex DL models, aiding in feature learning. DBNs are a variant of RBMs, functioning as generative models primarily used for feature learning. GNNs are designed for graph-structured data, empowering tasks like node classification and link prediction.
\item \textbf{Datasets:} Diverse benchmark datasets are available for each data modality, with varying characteristics and complexities.
\end{itemize}

\section{{Cover}-type-based deep steganalysis}
\label{sec4}

\ac{DL} techniques for steganalysis vary based on the cover type and data being analyzed, as different types of data have different statistical properties, and steganography methods use different techniques to embed secret information. For instance, image steganography may use \ac{LSB} embedding or spatial domain techniques, while speech steganography may use spread spectrum or echo hiding. Therefore, DL techniques need to be adapted and optimized for each type of data and cover type to effectively detect hidden information. The upcoming subsections provide a summary of the \ac{SOTA} methods proposed for each data type. Table \ref{TCoverBased} summarizes many of the detailed DL-based steganalysis schemes for both image and video data type.

\subsection{Image steganalysis}
This section provides an overview of the latest and most effective image steganalysis schemes that utilize \ac{DL} techniques. To assist researchers who may want to replicate or build upon these models, Figure \ref{CnnArchi} presents a comprehensive summary of the architectures of the most commonly used models, including detailed information, such as kernels, max-pooling, \ac{BN} layer sizes, among others. Theoretical definitions of image-based steganalysis, including feature-based steganalysis, low probability modification, primary and secondary steganalytic classifiers, and more, can be found in \cite{megias2022subsequent}. Additionally, numerous proposed DL-based steganalysis frameworks benchmark their performance against the \ac{SRM} method \cite{fridrich2012rich}. This framework is tailored for building steganography detectors by assembling a comprehensive noise model from various sub-models using image noise residuals. The model assembly is a key part of the training process, guided by samples from cover and stego sources. Ensemble \ac{ML} classifiers are used due to their efficiency in handling high-dimensional feature spaces and large datasets, given their low computational complexity.


\subsubsection{CNN-based {methods}}
Since 2015, the use of CNN-based steganalysis methods has been increasing, gradually replacing the two-step machine-learning-based steganalysis approaches that involve feature extraction and classification. This shift in preference is largely due to the improved performance offered by CNN-based methods. The following summarizes and clusters proposed CNN \ac{SOTA} methods that successfully identify stego images and clear cover images. \\

\begin{figure}[h!]
    \centering
    \includegraphics[scale=1.4]{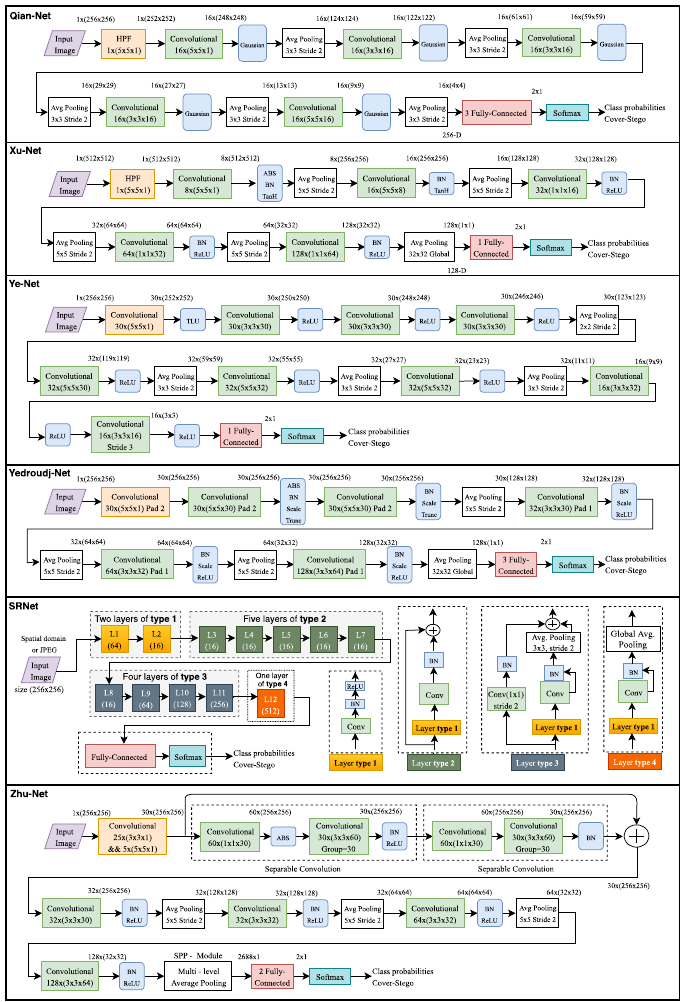}
    \caption{Commonly used CNNs schemes for the steganalysis of digital images. The information enclosed within the boxes contains the number of kernels multiplied by the height, width, and number of feature maps as input. On the other hand, the data outside the boxes is structured as the number of feature maps multiplied by height and width. If the values of Stride or Padding are not given, then the assumed values are 1 and 0, respectively \cite{reinel2019deep,tabares2020digital}. }
    \label{CnnArchi}
\end{figure}

\begin{figure}[h!]
    \centering
    \includegraphics[scale=1.2]{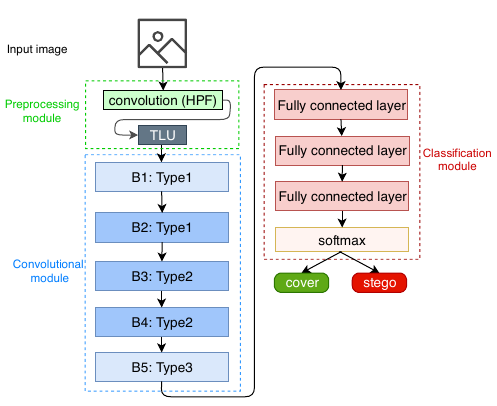}
    \caption{General outline for a steganalyzer that includes a preprocessing module besides a convolutional module for feature extraction, and a classification module.}
    \label{TLU}
\end{figure}

\noindent\textbf{(a) Pre-processing-based {methods}:}
\ac{CNN} based steganalysis is promising but suffers from degraded detection performance when an image is resized by the nearest-neighbor interpolation before steganography. To address this shortcoming, the preprocessing proposed in \cite{kato2020preprocessing} involves the additional embedding of steganographic signals into both resized original images and resized steganographic ones with the same embedding key. This helps \ac{CNN} learn features by making the difference of spatial frequencies between them obvious. In 2017, Ye et al. \cite{ye2017deep} proposed a novel approach that makes use of a set of high-pass filters in a \ac{SRM} to identify the steganographic signal in an image. They found that initializing the parameters of the preprocessing layer significantly improved the results compared to random initialization. Additionally, they introduced the use of a \ac{TLU} for the first time and designed a TLU-CNN based on this activation function, as depicted in  Figure \ref{TLU}. They also proposed the idea of selection-channel-aware steganalysis and developed a selection-channel-aware TLU-CNN network. The experimental results show that this network outperforms the traditional rich-model method in terms of detection performance. Yedroudj et al. \cite{yedroudj2018yedroudj} proposed a network called Yedroudj-Net that combines the successful features of  Xu-Net \cite{xu2016structural} and Ye-Net \cite{ye2017deep} as depicted in Figure \ref{CnnArchi}. They used 30 filters from SRM as initialization values for the preprocessing layer and added \ac{BN} layers and \ac{TLU}. Even without using selection-channel awareness, the network achieved good performance. Similarly,  \cite{lu2019importance} proposes a study on the use of a \ac{TLU} function in the pre-processing phase of a CNN-based steganalyzer. This function, combined with high-pass filters, can make the steganalyzer suitable for both spatial and JPEG domain steganalysis. This also allows the steganalysis method to perform detection in other domains, for instance, using Yedroudj-Net for JPEG steganalysis.  

\afterpage{\clearpage}

The performance of CNN-based steganalysis approaches is often linked to the size of the learning database. Increasing the size of the database can lead to better results, but obtaining a large database with controlled acquisition conditions can be difficult. Data augmentation can be an efficient approach to improve the efficiency of the steganalysis process. The proposed method in \cite{yedroudj2020pixels}, is called ``pixels-off'', which involves switching off a small proportion of pixels from the initial database of cover images to increase the data set. This technique is different from traditional noise addition methods, which alter the initial surface statistical regularities, leading to a mismatch between the learning set and the test set. The pixels-off technique is applied directly to the spatial domain, keeping the average cover statistics. This technique can simulate dead pixels or faulty sensors that frequently occur in various application domains.\\

\noindent\textbf{(b) Universal methods:} \ac{DL} has gained popularity in various fields and has also been applied to steganalysis by researchers, resulting in significant achievements. Steganalysis using \ac{DL} has been proposed by several researchers. Qian et al. \cite{qian2015deep} proposed a \acp{CNN} model that can automatically learn feature representations and complex dependencies that are useful for steganalysis. Zeng et al. proposed a hybrid deep-learning framework for the steganalysis of JPEG images that incorporates domain knowledge from rich models \cite{pibre2015deep}. Xu et al. \cite{xu2016structural}  introduced a \ac{CNN} structure with five convolutional layers and incorporated \ac{BN} and global average pooling, which are widely used in image classification tasks. The network applies several activation functions, including absolute activation, \ac{TanH} activation function, and \ac{ReLU} to improve the experimental results. Their performance surpasses the \ac{SRM} scheme \cite{fridrich2012rich}, and the improved Xu-Net yields better outcomes for steganalysis in the \ac{JPEG} domain \cite{xu2017deep}.  One of the classical methods includes the \ac{GNCNN} proposed by Qian et al. \cite{qian2016learning,qian2018feature}, which uses a Gaussian activation function with \acp{CNN}. The network structure is comprised of three parts: a preprocessing layer that uses high-pass filters, a convolutional layer for feature extraction, and a fully connected layer for classification. This method is one of the earliest to use \acp{CNN} for steganalysis and has produced results that are similar to traditional methods that use manually designed features.

\begin{figure}[h!]
    \centering
    \includegraphics[scale=1]{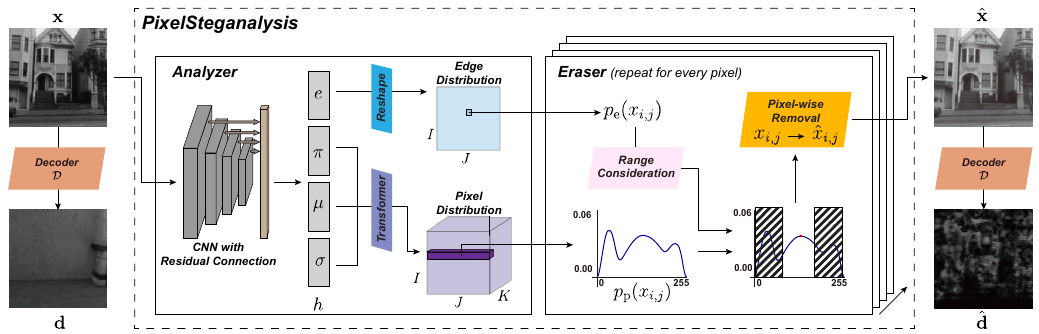}
    \caption{An example of pixel-level steganalysis. The PixelSteganalysis $P$ framework has two components, an analyzer, and an eraser. When given a stego image $x$ as input, the framework applies the eraser function to produce a purified stego image $\hat{x}$, denoted as $P(x)$.}
    \label{pixelSteg}
\end{figure}

Li et al. \cite{li2018rest} developed a \ac{CNN} architecture, called ReSTNet, that consists of both linear and nonlinear filters arranged in a parallel subnet structure, leading to enhanced steganographic detection performance.  Zhang et al. \cite{zhang2018efficient} proposed a \ac{CNN} model that utilizes separable convolution, multilevel pooling, and spatial pyramid pooling techniques to achieve high detection accuracy for images of various sizes.   Yang et al. \cite{yang2017steganalysis} used maxCNN to integrate knowledge from the selection channel into a \ac{CNN} for steganalysis. Zhang et al. \cite{zhang2022ag} proposed an improved general \ac{CNN} model called AG-Net for low-bit embedding steganalysis in the framework of \ac{DL}. The model includes a confrontation module to extract and compare features from cover and stego parts, an association between adjacent confrontation modules to accumulate feature differences and a softmax layer for classification and detection of stego images. The work in \cite{yang2018analysis} analyzes the impact of input data sequence on \ac{CNN} training results based on the training mechanism. Differences in the data sequence can be caused by data correlation, data mini-batch processing, and high-pass filtering operations. The results show that the pairwise data, comprising cover and stego images generated using two separate steganography algorithms (S\_UNIWARD and WOW) at distinct embedding rates (0.1 and 0.4), has a better network detection effect than the random data. Due to overfitting and hyper-parameter tuning problems, \ac{DL} models encounter difficulties.  Thus, in \cite{iskanderani2021artificial}, an effective \ac{CNN} model, denoted as $\theta$ \ac{NSGA-III}, is introduced. This model is built upon the \ac{DCNN} model and is employed to address these issues in image steganalysis by fine-tuning the initial model parameters. The universal method in \cite{su2019boosting}   includes an ensemble classification strategy to allow any CNN-based steganalysis to train just one model, forming an ensemble that can enhance detection accuracy for both spatial and JPEG steganography. The proposed method constructs sub-spaces for training base learners, and a novel voting fusion structure automatically optimizes during the training process.  Mustafa et al. \cite{mustafa2020enhancing} propose a new approach to blind image steganalysis that reduces the computational cost by discarding pre-processing, improves detection accuracy by modifying a recent \ac{CNN} model, and uses a hybrid technique of model and data parallelism in both convolution and fully connected layers. The training workflow is based on forward and backward propagation. The study \cite{liu2017ensemble} compared \ac{CNN} and the latest version of \ac{SRM-EC}, called maxSRMd2 method \cite{denemark2014selection}, and proposed an ensemble method that combined the two, yielding an improvement that surpassed the performance achieved by maxSRMd2.  The proposed framework \cite{jung2021pixelsteganalysis} aims to prevent the use of DL-based steganography in covert communications and transactions. The framework proposes a DL-based steganalysis technique that removes secret images by restoring the original image's distribution. The technique uses a \ac{DNN} to exploit pixel and edge distributions and remove the hidden information, contrary to the attention-based schemes which use a whole region of interest during classification. The evaluation of the technique using three public benchmarks shows better performance compared to conventional steganalysis methods, considering that the decoding method of DL-based steganography is different from conventional steganography. Figure \ref{pixelSteg} illustrates an example of a pixel-level steganalysis scheme.

WISERNet is a highly regarded color image steganalysis algorithm proposed in 2019 \cite{zeng2019wisernet}. The proposed method is based on using separate channel-wise convolution without summation in the bottom layer and united normal convolution in the upper layers. Thus, WISERNet widens the upper convolutional layer to increase the output involved in the summation and suppresses the image content in the lower layer, which improves network performance. The authors conducted extensive experiments and found that WISERNet outperformed other \ac{SOTA} steganalytic models for color images with less complexity. Ren et al. \cite{ren2020learning} proposed an end-to-end steganalytic scheme that involves a selection channel network and a steganalysis network. The two networks, LSCA-XuNet and LSCA-YeNet, are trained together, with the selection channel network learning and outputting selection channels for the steganalysis network.  Gomis et al. \cite{gomis2020estimation} focused on steganalysis of images with a cover-source mismatch problem. The study suggests using a combination of unsupervised and supervised \ac{ML} algorithms to improve the classifier's performance. The unsupervised step involves using the $k$-means algorithm to group similar images. When the number of extracted features is large, the study recommends using a \ac{CNN} to group similar images and an \ac{MLP} neural network to estimate the hidden message length in different groups. Lai et al. \cite{lai2022generative} made significant contributions to steganalysis by creating a U-shaped symmetric architecture that preserves spatial distribution properties, developing a focus enhancement filter module to improve the learning of low-amplitude hidden signals, using feedback residual and pixel shuffle techniques for high-quality image reconstruction, and conducting experiments that show the effectiveness of their proposed FFR-Net method compared to state-of-the-art steganalysis models. You et al. in \cite{you2019restegnet} proposed a new steganalytic network, called RestegNet, that uses a novel building block group consisting of sharpening and smoothing blocks, named ShRC and SmRC, respectively, to widen the gap between cover and stego images in feature extraction layers. The network outperforms previous methods obtained for the BOSSbase 1.01 data set. RestegNet achieves better detection accuracy and faster convergence speed compared to XuNet and TLU-CNN, while using the same preprocessing layers as the previous methods. \\

\noindent\textbf{(c) Attention layer-based {methods}:}
The attention mechanism is a technique that helps to prioritize important regions and generate feature weights efficiently. It was inspired by the human visual system, which selectively focuses on important information while ignoring less relevant details. By simulating this process, the attention mechanism can enhance the performance of classification networks by improving the focus on relevant information. The integration of attention and artificial intelligence tasks has become increasingly important for capturing global dependencies. Self-attention, which was developed later, allows for a response from a specific position in a sequence by focusing on all positions in that same sequence. This approach overcomes the limitations of convolution, which only operates on local neighborhoods and lacks global information. The SAANet network model proposed in \cite{huang2022image} replaces traditional convolution with attention-augmented convolution, which enables the network to allocate more learning weights to the steganographic area. This approach guides the network to learn features that are more advantageous for steganalysis.  Yang et al. \cite{yang2022multi} present a new image steganalysis model that uses a multilayer attention mechanism and a dual-residual structure network to enhance the transmission of steganographic signals. The proposed model is able to effectively capture steganographic data, as demonstrated by its performance on many data sets, achieving comparable results compared to current algorithms. The method proposed in \cite{liu2023image} uses a \ac{CNN} model with an efficient channel attention module by merging the \ac{ECA} module with \ac{CNN}, to focus on the steganographic region of the image and capture local cross-channel interaction information. \Ac{DTL} is applied to improve the steganalysis performance of low embedding rate images. Experimental results show that the proposed method outperforms the typical state-of-the-art models, demonstrating higher detection performance for steganalysis of low embedding rate images. The performance of \ac{DL}-based steganalysis can be enhanced by incorporating \ac{SCA} knowledge, but this is often unavailable in practical applications. To address this, the framework presented in \cite{liu2022effective} proposes a novel \ac{CNN} model that includes extra non-linear kernels in the first network layer to enhance the stego signal and utilizes the residual \ac{CSA} module along with a \ac{SPP} to improve performance, similar to \ac{SCA}.\\

\noindent\textbf{(d) Noise-based {methods}:} According to current literature, the steganalytic detector's accuracy improves if it is trained using the residual error (embedding noise) domain. However, to obtain an accurate noise residual, it is necessary to predict the cover image precisely from the corresponding stego image. In the scheme proposed in \cite{singh2021steganalysis}, a denoising kernel was used to obtain a more precise noise residual. Subsequently, a CNN-based steganalytic detector was developed, which was trained using the noise residual to improve detection accuracy. The results of the experiments indicate that the proposed approach outperforms current steganalysis schemes against current steganographic techniques. Boroumand et al. \cite{boroumand2018deepieee,boroumand2018deep} introduced SRNet, which does not use conventional high-pass filters but instead focuses on maximizing the noise residuals introduced by steganography methods, leading to one of the most accurate steganalysis methods to date. You et al. \cite{you2020siamese} designed an end-to-end \ac{DL} solution for detecting steganography images from normal images, without retraining parameters, using a Ke-Net architecture based on the Siamese network with three phases: preprocessing, feature extraction, and fusion/classification. The algorithm is validated using data sets of steganography images of varying sizes and their corresponding normal images, sourced from BOSSbase 1.01 and ALASKA \#2. The experimental results indicate that the proposed network is well-generalized and robust.

The scheme proposed in \cite{zhong2020deep} constructs and applies models for removing document image hidden messages using \ac{DL}, based on the concept of adversarial perturbation removal. These models include feed-forward denoising \ac{CNN} and \ac{HGD}. The method demonstrated better scalability and adaptability compared to \ac{SOTA} methods. The method in \cite{arivazhagan2021hybrid} proposes a hybrid \ac{DL} framework with a \ac{CNN} that uses residual-based investigation for digital image steganalysis. The proposed framework extracts five types of handcrafted features to model the noise residuals, and a robust  \ac{CNN} with a new parallel architecture is used to classify the input features as cover or stego. The \ac{CNN} is trained using stacked 1-dimensional residual features arranged in a 2-dimensional grid. The proposed architecture is evaluated using three spatial content-adaptive and four spatial non-content-adaptive algorithms, and the results show that it outperforms existing \ac{SOTA} steganalytic methods. The approach presented in \cite{lin2021multi} proposes a \ac{MFRNet} for the steganalysis of color images. \ac{MFRNet} uses \ac{DL} techniques to analyze color images and detect steganographic noise. The network consists of two modules: a color image preprocessing module and a multi-frequency steganographic noise residual extraction to detect high-frequency information in the image texture, as well as low-frequency information that corresponds to the image content, and classification module. The MFRNet is more lightweight than other \ac{DL} models used in steganalysis, making it more efficient for real-world applications. This research has potential applications in cybersecurity and digital forensics. \ac{MFRNet} can suppress the interference of image content and better reduce the impact of the steganography algorithm and payload mismatch. The experimental results show that \ac{MFRNet} outperforms the \ac{SOTA} model, WISERNet \cite{zeng2019wisernet},  on the PPG-LIRMMCOLOR data set.\\

\noindent\textbf{(e) Transform-based {methods}:} Due to the widespread use of \ac{JPEG} compression, several steganographic techniques have been developed to hide messages in \ac{JPEG}  format. Therefore, it is crucial to conduct steganalysis on JPEG images. Zeng et al. \cite{zeng2017large} were the first to use a \ac{DL} approach in the transform domain steganalysis, in 2016, and they introduced a new model called hybrid \ac{CNN} (HCNN) consisting of three \ac{CNN} subnetworks for JPEG steganalysis. The experimental results showed that HCNN outperformed other existing \ac{SOTA} models in terms of accuracy. However, compared to previous \ac{DL}-based steganalysis models in the spatial domain, HCNN is more complex. This is due to two additional steps, namely quantitative and truncated, and three paths in the feature extraction module. Previous models only had one path, and, consequently, HCNN leads to an increase in network complexity and computational overhead. FractalNet \cite{larsson2016fractalnet, gan2018efficient}, a \ac{DL} model, has recently demonstrated competitive performance similar to that of Resnet.

Reference \cite{gan2018efficient} presents a new CNN-based steganalysis model for JPEG images called JPEGCNN. It calculates pixel residuals using a 3$\times$3 kernel function, which solves problems with image content affecting direct analysis of \ac{DCT} coefficients and larger kernel functions not effectively capturing neighborhood correlation changes. Compared to the HCNN model, JPEGCNN is a lightweight structure with fewer training parameters and maintains similar accuracy. The OneHot method in \cite{yousfi2020intriguing}, for JPEG steganalysis, uses DCT coefficients as a preprocessing step, and demonstrated that a \ac{DNN} may not be necessary to achieve desired accuracy levels. Only a small number of layers were found to be sufficient. The work proposed in \cite{chen2017jpeg} presents a method for adapting CNN-based detectors for detecting modern JPEG-domain steganography. The method involves introducing the concept of JPEG phase awareness into the network architecture through a new phase-split layer and two ways for incorporating phase awareness. The authors also enhance the key-value kernel, a kernel size traditionally used to pre-filter images for \ac{CNN} detectors, with a second fixed kernel to enable the network to learn kernels that are better suited for detecting the stego noise introduced by JPEG-domain embedding techniques. The proposed design is tested using J-UNIWARD \cite{su2020fast} and UED-JC \cite{guo2014uniform} steganographic algorithms in experiments to show its effectiveness. The work presented in \cite{chen2018deep} proposed quantitative steganalysis that aims to estimate the size of the hidden message within an image that has been identified as containing confidential data. The features extracted from the activation of \ac{CNN} detectors are used to perform payload estimation in the spatial and \ac{JPEG} domains.  The authors refer to this method as the "bucket regressor," which utilizes the \ac{MSE} activation function for prediction instead of softmax for classification. Zhan et al. \cite{zhan2021advanced}  proposed a JPEG steganalysis method based on FractalNet, called FRNet. This method involves introducing a residual unit with a shortcut connection into the fractal structure, which enables the network to effectively suppress image content and generate the residual image with stego noise. A deep feature extraction module is then used to downsample the feature map and superimpose the weak stego signal between different channels of the convolution layer. The proposed approach achieved good detection performance when evaluated on two steganographic algorithms, namely, J-UNIWARD and UERD.\\

\begin{figure}[h!]
    \centering
    \includegraphics[scale=1.2]{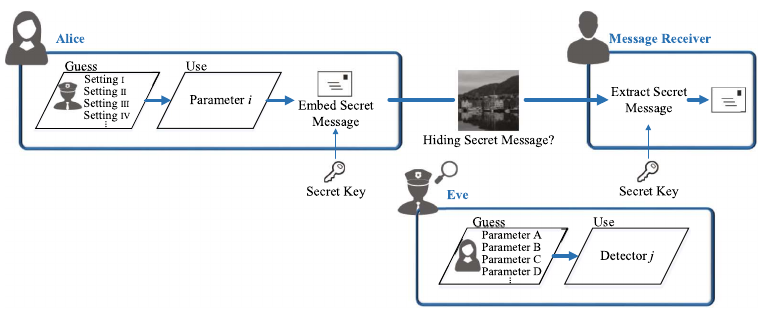}
    \caption{ The issue of embedding or detecting steganography in an adversary-aware context.}
    \label{advSteg}
\end{figure}

\noindent\textbf{(f) Adversarial-based {methods}:} \ac{CNN}-based steganalysis has demonstrated superior performance compared to traditional \ac{ML} methods. However, recent research in \ac{DL} has revealed that \acp{CNN} are susceptible to adversarial examples, which are slight modifications of input data that can cause incorrect decisions. Adversarial examples have been used successfully in steganography to counter target \ac{CNN} steganalyzers by adjusting the embedding costs based on back-propagated gradients \cite{tang2019cnn}. Adversary-aware training can improve the performance of the steganalyzer, but this can lead to a dilemma where Eve, the adversary, does not know the exact parameters used by Alice, the steganographer, for adversarial embedding \cite{barni2018adversarial}. Similarly, Alice cannot tune the internal parameters of the steganographic scheme to maximize its deception capability because she does not know the exact images used by Eve to train the steganalyzer. This dilemma is illustrated in Figure \ref{advSteg}. To overcome the apparent stalemate, Shi et al. \cite{shi2020cnn} propose a game theoretical approach to address the problem of determining the parameters for the steganalyst and steganographer. They suggest two variations of the game, differing in how the steganalyzer-based \ac{CNN} network output is thresholded to reach a final decision. In both games, the steganographer aims to increase the probability of missed detection, while the steganalyst aims to reduce the overall error probability in one game and missed detection probability for a given false alarm rate in the other. They conducted experiments using a modified version of the adversarial embedding scheme proposed by Tang et al. \cite{tang2019cnn} and found equilibrium points and corresponding payoffs for both versions of the game. By comparing error probabilities at equilibrium to those obtained using alternative strategies, such as a worst-case assumption or Tang et al.'s adversarial embedding scheme, they demonstrate the advantages of using game theory to address the interaction between the steganographer and steganalyst.

\begin{figure}[h!]
    \centering
    \includegraphics[scale=1]{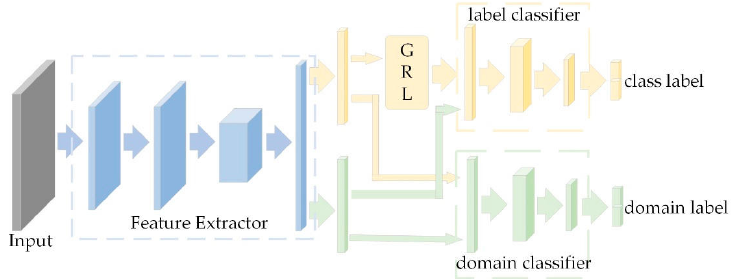}
    \caption{ An example of adversarial steganography using \ac{GRL}.}
    \label{advGRL}
\end{figure}

The purpose of the scheme proposed in \cite{ge2021novel} is to suggest a novel end-to-end network that can enhance steganalysis task performance by balancing detection accuracy and efficiency. The authors have incorporated separable convolution and an adversarial mechanism to isolate the steganographic from the content signals in spatial images. This separation enables better extraction of steganographic embedding features and enhances the performance of image steganographic detection by reducing interference from the image content. The \ac{GRL} makes adversarial training between the feature extractor and the label classifier more straightforward. \ac{GRL} optimizes the loss function in the opposite direction of the gradient, resulting in the extraction of more image content features to deceive the discriminator's classification. Figure \ref{advGRL} illustrates the scheme proposed in \cite{ge2021novel}. Zhang et al. \cite{zhang2022adnet} suggest a \ac{CNN} model for detecting adversarial examples, which is based on end-to-end based- attention steganalysis. The work proposed in \cite{zhang2019enhancing}  presents evidence that training the YeNet architecture \cite{ye2017deep}, which can undergo training using paired data sets of both cover images and corresponding stego images, with adversarially generated examples from the SteganoGAN method \cite{zhang2019steganogan}, can lead to improved performance on the blind steganalysis task.

\subsubsection{DRN/GNN-based {methods} }

\Acp{DRN}, or ResNets, were proposed initially by He et al. \cite{he2016deep} for image recognition. The  success of \ac{DRN}  in tasks such as image classification, object detection, and semantic segmentation has led to its use in the field of steganalysis in numerous works 
\cite{xu2017deep,yousfi2019breaking,boroumand2018deepieee,li2019non,tan2020calpa,chen2022deep,wu2018deep,boroumand2018deep,zhang2019deep,zeng2020deep,huang2020deep,xu2020drhnet,li2020ensemble,lee2020deep, wang2021steganalysis,kang2021identification,chen2022image,singhal2021multi,wu2017residual,wu2016steganalysis,ozcan2018transfer,zhang2022generative}. Wu et al. were the first to apply a \ac{DRN} in image steganalysis in their works  \cite{wu2016steganalysis, wu2018deep, wu2017residual}. Their proposed architecture includes three sub-networks: a \ac{HPF} sub-network, a \ac{DRL} sub-network, and a classification sub-network. The \ac{HPF} sub-network extracts noise components from cover and stego images. The \ac{DRL} sub-network extracts relevant features to distinguish between cover and stego images. The classification sub-network, a fully connected neural network, maps the features obtained from the Residual Learning sub-network to binary labels. Tan et al. \cite{tan2020calpa} and Li et al. \cite{li2019non} proposed incorporating channel pruning to shrink the size of the steganalyzer while preserving accuracy. They applied the pruning method to both the SRNet and XuNet2 architectures (Figure \ref{CnnArchi}). The experimental results show that the proposed non-structured pruning method can significantly reduce the computational and storage cost of the original deep-learning frameworks without affecting their detection accuracy.

The SRNet model, illustrated in Figure \ref{CnnArchi}, is a \ac{DRN}-based steganalysis model that has been used as a foundation for multiple works, which have then been improved upon \cite{yousfi2019breaking, li2020ensemble, wang2021steganalysis, huang2020deep,zhang2022dataset,chen2022deep,chen2022image,zhang2022generative,lee2020deep}. For example, Yousfi et al. \cite{yousfi2019breaking} presented the design and training of detectors developed for the ALASKA steganalysis challenge. The detectors, based on SRNet, were trained on different combinations of three input channels: luminance and two chrominance channels, to handle varying image quality factors. This approach is based on the idea that the different channels in an image have distinct statistical properties that can be utilized to enhance steganalysis performance. On the other hand, Chen and Jiao \cite{chen2022deep} introduced a new neural network called ``SRNet-CBAM'' that incorporates a~\ac{CBAM} to address the challenges of analyzing adaptive steganography and identifying picture regions that are favorable for modeling.  This mechanism helps to identify and highlight important features of an image, such as high-frequency information while ignoring less relevant details like smooth background. As a result, the network can better represent the embedded signal. The experimental results show that the SRNet-CBAM model improves the accuracy of the original SRNet model by an average of 1.36\%.


Li et al. \cite{li2020ensemble} introduced a method for maximizing the accuracy of steganalysis through ensemble learning. This method is built on the SRNet model and incorporates multiple base learners that are derived from different training data and snapshots. The optimization is achieved through a combination of decision outputs and feature fusion. The results of their experiments demonstrate that this ensemble method improves classification accuracy compared to the SRNet model in both the spatial and JPEG domains. 
Wang et al. \cite{wang2021steganalysis} applied \ac{DRN} based on \ac{NAS} to enhance the SRNet by discovering a combination of residual layer structures that is more appropriate for steganalysis. Additionally, they designed a long-span residual structure that preserves the signal of the hidden message without weakening it after passing through a multi-layer \ac{CNN}. The findings demonstrate that this model outperforms manually designed models in most steganography processing.   In \cite{xu2020drhnet}, the authors present DRHNet, a model for the steganalysis of images that employs heterogeneous kernel residual learning. Rather than using the image directly as input to the network, they extract features and create a feature matrix with a rich model, which serves as the actual input. They also use a heterogeneous kernel called ``HetConv'', leading to reduced computational complexity without sacrificing accuracy. In \cite{singhal2021multi,kang2021identification}, \ac{DRN}  has been used for multi-class blind image steganalysis. This approach identifies the stego image into the corresponding embedding algorithm class without any prior knowledge of the cover media or steganographic algorithm used. However, \cite{kang2021identification} employed hierarchical \acp{DRN}.

\Acp{GNN} have been also exploited in image steganalysis. Liu and Wu \cite{liu2022graph} claim that altering an image for the purpose of steganography inevitably impacts the statistical properties of the graph-based features that are extracted from the cover image. To address this, they convert each image into a graph, where the patches of the image are represented as nodes and the relationships between the patches are represented as edges. The features of each node are derived from its corresponding patch using a shallow \ac{CNN} structure. The graph is then processed through a graph attention network for feature representation, enabling the feature vector produced to be used for final classification. The results of their experiments show that the proposed architecture performs similarly to the commonly used \ac{CNN} model.

\subsubsection{AE/DNN-based {methods} }

Chen et al. \cite{chen2022image}, suggested a \ac{DNN} model with multiple scales based on the steganalysis residual network SRNet. They selected various local receptive fields within the same layer to produce distinctive channels of features. These channels were then recognized to identify different steganographic features of the image at different scales. Their experiments demonstrated that using a multi-scale residual network resulted in higher accuracy in detecting steganography compared to networks that only used a single-scale channel.

\begin{figure}[bt!]
    \centering
    \includegraphics[scale=1]{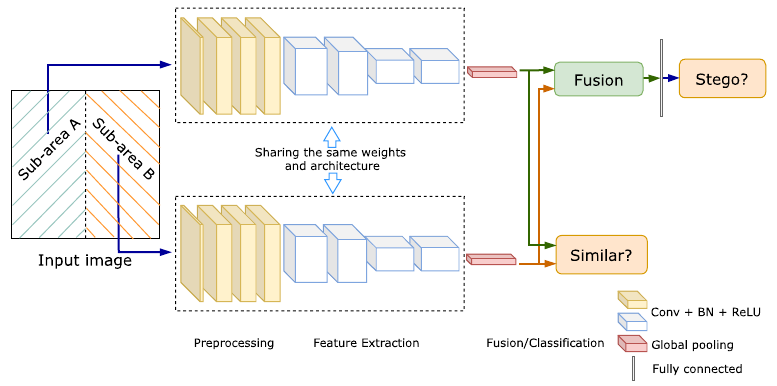}
    \caption{An example of using a Siamese architecture that relies on two sub-areas of an input image or composite steganography/non-steganography image that are separately processed by parallel subnetworks. Noise residuals are then extracted from each sub-area, and the outputs of the two subnetworks are combined to extract relationships that distinguish steganographic images from non-steganographic ones. The Siamese architecture has been used in \cite{you2020siamese}.}
    \label{Siamese}
\end{figure}

In 2014, Tan and Li \cite{tan2014stacked} introduced an approach that incorporated \ac{AE} for detecting steganography in images, which involved stacking multiple layers of convolutional auto-encoders. However, despite being able to detect steganography in various image formats and not just JPEG images, this method failed to surpass the performance of pre-existing steganalysis techniques. Cohen et al. \cite{cohen2020assaf} proposed an architecture that consists of two stages. In the first stage, a \ac{DAE} is applied to perform preprocessing on the input image. In the second stage, a Siamese neural network classifier is proposed. This latter network measures the distance between the DAE-processed image and the original input image to make this determination (Figure \ref{Siamese}). The idea behind this architecture is that the \ac{DAE} can learn the preprocessing steps for the image during training. 

\begin{center}
\scriptsize
\begin{longtable}[!t]{m{0.5cm}m{1cm}cm{3.3cm}m{1.3cm}m{1.5cm}m{1.2cm}m{5cm}}
\caption{Cover-based image/video steganalysis.  The type of cover could be identified from the data set used. (S), (T), and (S-Temp) stand for spatial, transform, and spatial-temporal domains, respectively. The well-known steganographic schemes: WOW, S-UNIWARD, HILL, and MiPOD belong to the S domain. UED, UERD, and J-UNIWARD, belong to the T domain. When many results are presented, only the maximum performance is mentioned. The improvement is calculated in percentage points.} \label{TCoverBased}\\
\hline
 Paper& DL model& Domain & Steganographic algorithm & Compared to &Data set &Metric used  & Observations\\ \hline
\endfirsthead
\multicolumn{6}{c}{Table \thetable\ (Continue)} \\
\hline
 Paper& DL model& Domain & Steganographic algorithm & Compared to &Data set &  Metric used & Comments\\ \hline
\endhead
\hline
\endfoot
\hline \hline
\endlastfoot
\cite{kato2020preprocessing} &CNN& S & WOW, S-UNIWARD, HUGO-BD & \cite{boroumand2018deepieee} & BOSSbase 
 & Accuracy &Detection improved up to 34.8\% when the same steganogram is  embedded.\\
 
\cite{ye2017deep}&  TLU-CNN &  S & WOW,  S-UNIWARD,  HILL &  \cite{fridrich2012rich,denemark2014selection} & BOSS, BOWS2& DER & A novel activation function \ac{TLU} used while integrating channel selection knowledge to enhance its performance. \ac{DER} ranges from  11.82\% to 28.08\% depending on dataset preprocessing. \\ [0.3cm]

\cite{lu2019importance} &YedroudjNet, &S, T&WOW, J-UNIWARD& \cite{yedroudj2018yedroudj,yang2017jpeg}&BOSSBase & Accuracy & DenseNet is also investigated.  In only 100 epoch, accuracy up to 90\%.\\
 \cite{qian2015deep}& GNCNN & S & HUGO, WOW, S-UNIWARD & \cite{fridrich2012rich} & BOSSbase, ImageNet& DER & Improvement from 2\% to 5\% higher \ac{DER} than \ac{SRM} method.\\
 
\cite{pibre2015deep} & CNN & S & HUGO, WOW, S-UNIWARD & \cite{qian2015deep} & BOSSbase & DER & Achieving a 17\% reduction in \ac{DER} compared to the SRM-EC method.\\

\cite{xu2016structural}& CNN & S & S-UNIWARD, HILL & \cite{fridrich2012rich} & BOSSbase & Accuracy, ROC & Achieving accuracy and \ac{ROC} results comparable to that obtained using the SRM method.\\

\cite{xu2017deep}&CNN&S& S-UNIWARD&\cite{fridrich2012rich,denemark2014selection} &BOSSbase & DER &Integrating the networks from \cite{ye2017deep} and \cite{he2016deep}.  Reduce \ac{DER} by about 35\%.\\

\cite{qian2018feature}  &  CNN  &  S  &  HUGO, WOW, S-UNIWARD, MiPOD, HILL-CMD  &  \cite{fridrich2012rich,denemark2014selection}  &   BOSSbase & DER  &  Outperform \ac{SOTA} methods in terms of \ac{DER} with 16.55\% for HOGO and 24.20\% for S-UNIWARD methods.  \\

 \cite{yang2018analysis} &  CNN  &  S  &  WOW, S-UNIWARD  &  Random data &  BOSSbase & Accuracy  &  The performance of in-pair data is better than that of shuffled data, with a maximum accuracy reaching 86\%. \\
 
 \cite{mustafa2020enhancing}&CNN& S &HILL, WOW, S-UNIWARD&\cite{yedroudj2018yedroudj,zhang2018efficient}&BOSSbase, ImageNet & Accuracy & Improvement reach 13.3\%, LReLU activation function is used. Speedup 21.2 times than \ac{SOTA}.\\
 
 \cite{jung2021pixelsteganalysis, }  &  CNN  &  S  &  LSB insertion, HUGO, WOW, S-UNIWARD  &  \cite{wang2004image}  &  Cifar-10, BOSSbase,  ImageNet & PSNR, SSIM, DT  &  Enhancements obtained in both the decoded rate and the \ac{DT}, ranging from 10\% to 20\%. \\
 
\cite{zeng2019wisernet} &  WISERNet  &  S  &  HILL, CMD-C-HILL, SUNIWARD, 
CMD-C-SUNIWARD  &  \cite{xu2016structural,ye2017deep,goljan2014rich}  &  BOSSBase & Accuracy, FLOPs  &  Enhanced detection with <50\% \acp{FLOP} complexity compared to advanced \ac{DL} colored image steganalyzer. \\

 \cite{gomis2020estimation} &  DeepStego &  T  &  Stegano   &  Universal  &  MNIST & Accuracy  &  MLP is used to estimate steganogram lengths with an accuracy reaching up to 99\%.  \\
 
 \cite{lai2022generative} &   FFR-Net  &  S, T  &  HILL, J-UNIWARD, WOW, UERD  & \cite{fridrich2012rich,denemark2014selection}  &  IStego100K, BOSSbase & MSE, PSNR, SSIM, DER, accuracy & Achieve cutting-edge results with accuracy of 97.04, 29.06, 0.88, and 229.38 for PSNR, mean SSIM, and MSE, respectively.   \\
 
 \cite{you2019restegnet}  & RestegNet & S & S-UNIWARD & \cite{xu2016structural,ye2017deep}  & BOSSbase & Accuracy  & RestegNet is improved by 3.07\% to 7.45\% compared to
XuNet, and faster 8 times compared to TLU-CNN. \\

 \cite{huang2022image} &  SAANet  &  T  &   J-UNIWARD, JC-UED  &  \cite{boroumand2018deep,chen2017jpeg,denemark2014selection,xu2017deep}  &  BOSSbase, BOWS2 & Accuracy &  Achieved up to 96.68\%. Model efficiency up to 28s. \\
 
 \cite{liu2023image} &  Attention &  S  &  S-UNIWARD, WOW, HUGO  &  \cite{xu2016structural,yedroudj2018yedroudj}  &  BOSSbase &  { Accuracy}  &  At an embedding rate of 0.05 \ac{bpp},  {the enhanced accuracy now spans from 5.03\% to 37.83\%.} \\
 
 \cite{boroumand2018deep} &  CNN  &  T  &  Low-pass and High-pass filtering, Denoising, Tonal  &  ML detector &  BOSSbase &  {Accuracy}  &   {Performance up to 97.91\%}. It can determine the processing history of JPEG images, it can be applied to both steganalysis and forensics.  \\
 
\cite{you2020siamese} & SiaStegNet  &  S  &  WOW, S-UNIWARD, HILL&  \cite{boroumand2018deepieee}  &  BOSSbase, \newline 
ALASKA \#2 &  {AUC}  &  {Achieved \ac{AUC} values exceeding 99\% in less than 100 epochs}, showcasing robust generalization abilities and can withstand variations.  \\

\cite{arivazhagan2021hybrid} & CNN & S & HUGO-BD, S-UNIWARD, WOW & \cite{fridrich2012rich,denemark2014selection} & BOSSbase  &  {Accuracy, DER} & Maximum accuracy of 83.6\% and 47.3\% in high and low volume payload bins, for content-adaptive and non-content-adaptive algorithms, respectively. \ac{DER} decrease up to 13.4\%.\\

\cite{zeng2017large}  &  CNN  &  T &  J-UNIWARD, UERD, UED  &  \cite{xu2016structural,xu2017deep}  &  ImageNet &  {FLOP}  &  Using quantization and truncation techniques in \ac{DL} steganalyzers can enhance their performance in terms of \acp{FLOP}.  \\

\cite{yousfi2020intriguing}  & OneHot CNN  &  T  &  J-UNIWARD, UED-JC, EBS, nsF5  &  \cite{boroumand2018deep,xu2017deep}  &  BOSSbase, BOWS2 \newline (Gray-scale) &  {DER} &   Able to learn relevant inter-block and intra-block connections using dilated convolutions and vanilla.  {4.7\% better than SRNet}.    \\

\cite{chen2017jpeg}  & PNet, VNet  &  T  & J-UNIWARD, UED-JC  &   SCA-GFR  &  BOSSbase, BOWS2 &  {Accuracy} &    {Improves detection by 7\%, while SCA-GFR only enhances it by 1\%.}  \\

\cite{zhang2022adnet} &  ADNet  &  S  &  C\&W, FGSM, BIM, DeepFool  &  \cite{ye2017deep,liu2019detection,liang2018detecting,xu2017feature}  &  ImageNet (ILSVRC-2016) &  {Accuracy} &   {Achieved 93.56\% for adversarial examples detection.} \\

\cite{zhang2019enhancing} &  YeNet  &  S  &  LSB, HiDDeN, SteganoGAN  &  ---  &  COCO &  {AUC} &  Able to detect alteration with an AUC of 97.10\%.  \\[0.2cm]

\cite{zhang2019deep} & DRN & S & HUGO, S-UNIWARD, WOW & \cite{denemark2014selection,xu2016structural,ye2017deep,qian2016learning} & BOSSbase, LIRMMBase &  {DER}  & When tested with a 0.4 payload, it showed a 1.2-3.5\% decrease in detection error across three data sets. \\[0.2cm]

\cite{zeng2020deep} & DRN & S & PDM, FDM, DHSPT, PTHVS & PMMTM,  LargeLTP & Created (IBBHalftone) &  {Filter size} & Apprach developed for halftone images.  {Filter size $5 \times 5$ can capture more embedding trace.}\\

\cite{huang2020deep} &  VSRNet  &  S  &  \Ac{CMV}   &  Aly, and Xu methods&  Xiph video test  media &  {MAE}  &  Videos are coded using HEVC.  {\ac{MAE} can be greater than 0.05. } \\

\cite{wang2021steganalysis} & SNNet  &  S  &  SUNIWARD, WOW   &  Change embedding rate  &  BOSSbase &  {AUC}  &   {Achieved up to 98.25\%, with an improvement of up to 1.5\%.} \\

\cite{kang2021identification}& ResNets& S & LSB, PVD, WOW, 
S-UNIWARD & -- & BOSSbase &  {Accuracy} & Attains a 79.71\% accuracy rate in classifying the four steganographic algorithms.  \\

\cite{zhang2022generative}& GLSNet & S & S-UNIWARD,  Baluja-Net \cite{baluja2017hiding} & Without RES, CNN, FC & BOSSbase &  {Accuracy} & Achieves a high accuracy of 77.75\% with only one pair of cover and stego images. \\

\cite{weng2022lightweight} &  LWENet  &  S  &  WOW, S-UNIWARD, HILL  &  \cite{yedroudj2018yedroudj,li2018rest,boroumand2018deep,deng2019fast,wang2021lightweight}  &  BOSSBase, BOSW2 &  {Accuracy} &Improving detection by using multi-view features, and   {achieved  92.59\%}.   \\

\cite{yang2017jpeg} & Xu-Net & T &J-UNIWARD& \cite{xu2017deep} & BOSSbase, BOWS2, ImageNet &  {DER} & The study found \ac{DER} of 40.26\% and 15.05\% for 0.4 and 0.1 embedding rates, respectively. \\

\cite{qin2022feature} &  CNN  &  S  &  CornerSearch, C\&W-$l_0$, SparseFool, DeepFool, C\&W-$l_2$, DDN, PGD, BIM  &  \cite{fridrich2012rich,liang2018detecting}  &  ImageNet, CIFAR-10 &  {Accuracy}  &  Steganalysis for adversarial examples detection.  {Achieving a result of 100\% for the CIFAR-10 dataset.}\\

\cite{zhang2022cnn}&CNN  &T  &DCT/DST &  SFE-AU, NRCNN  &  --- &  {Accuracy}
 & Based on the residual convolution layer for feature extraction,  {it achieved 91.81\%} in binary HEVC classification. \\
 
 \cite{yao2017deep} &  CNN  &  S-Temp &  Object forgery in video  &  \cite{chen2015automatic}  &  SYSU-OBJFORG &  {F1-score}  &   {Achieves up to 92.23\%} on SYSU-OBJFORG.  \\
 
\cite{9053969}&  SSTNet  &  S-Temp  &  AE-based DeepFakes, Face2Face, FaceSwap   &  FF++ &  FaceForensics++ &  {Accuracy} &   {Achieved up to 98.57\%}. The generalization  is proven on the GAN-based DeepFakes data set. \\[0.2cm]

\cite{yang2018deep} &  CNN  &  T  &  J-UNIWARD  &  \cite{xu2017deep}  &  BOSSbase, ImageNet &  {DER} &  Reduce \ac{DER} by 4.33\% and 6.55\%, for XuNet and SCA-GFR methods,
respectively.  \\[0.2cm]

\cite{prasad2020detection} &  CNN  &  S  &  HUGO, WOW, S-UNIWARD  &  --  &  BOSSbase, RAISE &  {Accuracy} & Steganalysis over noisy channel  {can reach an  accuracy of 97.5\%.} \\[0.2cm]

\cite{liu2022feature} &  FPNet  &  S  &  HiNet, WOW, S-UNIWARD, MiPOD, HUGO  &  \cite{ye2017deep,boroumand2018deepieee,you2020siamese}  &  COCO  &  {Accuracy} &  An ultra-lightweight network with just 0.16 million parameters, can achieve an accuracy of up to 91.75\%.  \\

\cite{ghosh2022deep} & DNN & S & StegHide & LSB, \cite{nissar2013texture} & LFW, BOSSbase &  {Accuracy} & Achieves  an accuracy of 90.93\% and 84.63\%  for LFW and BOSS data set respectively.  \\

\cite{plachta2022detection}  & DNN/ Ensemble classifiers & T & UERD, J-UNIWARD, nsF5 & \cite{wang2021steganalysis} & BOSSbase &  {Accuracy} & Achieves accuracy of 99.9\% for NsF5 (at 0.4 bpnzac), whereas for J-UNIWARD, detection was challenging at 0.1 bpnzac density (max. 56.3\%). Ensemble classifiers showed promise as an alternative to \ac{DL} detection. \\

\cite{bonnet2021forensics}& SRNet & S & HILL, MiPod & \cite{fridrich2012rich,goljan2014rich} &ImageNet &  {TPR} &  {Detection probability can reach 94.8\%}. The probability of fooling both  EfficientNet-b0 and the detector SRNet is 5.9\%. \\

\cite{zhang2020cover}& J-Net & S & S-UNIWARD, WOW & Without JMMD &BOSSBase &  {Accuracy} & Adapts JMMD into deep steganalysis.  {It can enhance the accuracy by 7\%-10\%}.  \\

\cite{deng2022universal} & CNN & S,  T & S-UNIWARD, HILL, WOW,  J-UNIWARD, UED-JC & SRNet, CovNet &BOSSBase, BOWS2 &  {Accuracy} &  {Attains a 0.56\% improvement compared to SRNet in the specific domain. }\\

\cite{zhang2019new}& CNN & T & J-UNIWARD, UED  & Ensemble classifiers &BOSSBase &  {Accuracy} & Outperforms traditional algorithms  by 2\%-3\% under different embedding rates. \\[0.3cm]

\cite{deng2021spatial} & PC-DARTS& S &S-UNIWARD, HILL, WOW &\cite{boroumand2018deepieee,ye2017deep} &BOSSbase, BOWS2&  {Accuracy} & Outperforms SRNet in detecting WOW at 0.4 bpp with 91.40\% accuracy, but shows slightly lower performance in other cases with a difference of less than 0.57\%.\\[0.4cm]

 \hline
\end{longtable}
\end{center}

\subsection{Video steganalysis}


Video steganalysis, unlike other data modalities such as images or text, has received comparatively less attention in the research community. This limited focus on video steganalysis can be attributed to several factors, including the complexity of video data, the increased computational demands of processing videos, and the challenges associated with developing steganalysis techniques specifically tailored to video content. Due to the constraints mentioned above, most steganalysis researchers opt to transform videos into sequences of images, enabling them to create steganalysis techniques applicable to both image and video formats (as depicted in Figure \ref{TaxonomySteg}). Additionally, numerous researchers have developed video steganalysis techniques and subsequently compared their outcomes with image steganalysis algorithms. The unequal distribution of research focus has led to a scarcity of well-established methodologies, benchmark datasets, and comprehensive studies in the field of video steganalysis. In this review, we have endeavored to present the available literature on video steganalysis to the best of our ability, considering the limitations in terms of the depth and breadth of research in this area.


To provide context, it's crucial to understand the pressing need for advancements in video steganalysis.

Advanced video encoding formats have become popular on the Internet, but this has led to increased video tampering and malicious video forgery. To address this issue, there is a growing need for passive video forensic techniques to verify the authenticity of pre-existing video content, which has become an important research topic. The researchers consider object-based forgery in video frames similar to image tampering in the corresponding motion residuals. They use feature extractors and algorithms that were initially designed for image steganalysis to extract forensic features from the motion residuals \cite{tan2022hybrid}. For example, reference \cite{yao2017deep} proposes a method for forgery detection that has two steps: video sequence preprocessing and network model training. In the first step, an absolute difference algorithm is used to convert the video sequence to image patches. These patches are labeled as positive and negative samples to create the training data set. In the second step, the data set is processed and used to train a five-layer CNN-based model. The proposed scheme can detect object-based steganographic methods in a video sequence \cite{bouzegza2022comprehensive}.


The use of \ac{DCT/DST} based steganography is a common practice and therefore, the corresponding steganalysis methods have become a relevant research topic. However, there are only a limited number of \ac{HEVC} video steganalysis methods that focus on the \ac{DCT/DST} domain, particularly motion vector and inter-block partition type \ac{HEVC} steganalysis.  The scheme in \cite{zhang2022cnn} presents a new steganalysis method for detecting \ac{DCT/DST} transform-based steganography in \ac{HEVC} video. The proposed method analyzes the distortion caused by \ac{DCT/DST}-based \ac{HEVC} steganography and its impact on pixel values, and uses a \ac{CNN} composed of residual convolution layers, steganalysis residual block modules, and a squeeze-and-excitation (SE) block for feature extraction and binary classification. Experimental results show that the proposed method outperforms existing steganalysis methods for detecting \ac{DCT/DST}-based \ac{HEVC} steganography. 
The work proposed in \cite{huang2020deep} focuses on quantitative steganalysis for videos, which is important for breaking secret codes in practical scenarios. Most video steganographical algorithms modify motion vector values in the compressed domain to embed secret messages. The work proposes a general framework for constructing video quantitative steganalyzers that use deep \ac{CNN} to detect motion vector embedding based on learned features, and discusses the construction of input data matrices for \ac{CNN} and the robustness of detection networks against different bitrates. Multiple models are used to extract features for constructing feature vectors. Experimental results show that the proposed method performs well, obtaining satisfactory estimation accuracy for testing \ac{HEVC} videos at multiple embedding rates under different video bitrates.

Huang et al. \cite{huang2020selection} presents a new approach to video steganalysis using a selection-channel-aware \ac{DNN}. Because videos have a different structure than images, the focus is on constructing an appropriate input data matrix for the deep \ac{CNN}, defining probability for motion vector modification, and using selection channel knowledge in the network structure. The proposed approach is inspired by SRNet, one of the most powerful image steganalysis neural networks, and is adapted for video steganalysis using a selection-channel-aware CNN-based scheme. The experimental results show that the proposed method performs well and benefits from selection channel knowledge when applied to testing \ac{HEVC} videos. Liu et al. \cite{liu2020steganalysis} designed a universal steganalysis method that is capable of detecting both intra-prediction mode and motion vector-based steganography using \ac{DL}. The proposed method, called noise residual \ac{CNN} (NR-CNN), operates in the spatial domain and integrates feature extraction and classification modules into a single trainable network framework. NR-CNN outperforms that of IS-Net and led to better steganalysis performance. In \cite{wu2020sstnet}, SSTNet framework is proposed for detecting manipulated faces incorporates three types of features: spatial, steganalysis, and temporal features. Spatial features are extracted by a \ac{DNN} for finding visible tampering traces like unnatural color, shape, and texture. Steganalysis features are extracted by a proposed novel constraint for suppressing high-level image content and detecting hidden tampering artifacts like abnormal statistical characteristics of image pixels. Temporal features are extracted by a recurrent network for discovering inconsistency between consecutive frames. 

\begin{figure}[h]
    \centering
\includegraphics[scale=1.1]{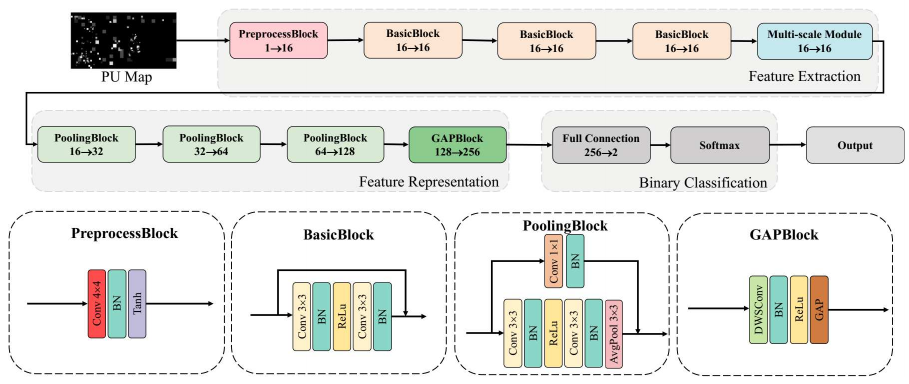}
    \caption{The overall framework of PUSN. For each block, $x_1\rightarrow x_2$ indicates that the input channel is $x_1$, and the output channel is $x_2$ \cite{dai2023hevc}. }
    \label{vidSteg}
\end{figure}

 {Current steganalysis techniques for \ac{PU}-based steganography mainly rely on extracting video statistical features, which overlook the potential information within each frame and struggle to accurately identify different \ac{PU}-based steganography methods. The study in \cite{dai2023hevc} presents a novel video steganalysis approach utilizing \ac{PU} maps and a multi-scale convolutional residual network. Initially, the impact of \ac{PU}-based steganography on both spatial and compressed domains is analyzed. It is observed that steganography minimally affects the spatial domain but significantly disrupts connections between \ac{PU} blocks in the compressed domain, creating distinct steganographic traces. Consequently, \ac{PU} partition modes with local connections are introduced to generate \ac{PU} maps for steganalysis. Subsequently, a video steganalysis network, termed prediction unit steganalysis network (PUSN) is developed. This network takes \ac{PU} maps as input and comprises three components: feature extraction, feature representation, and binary classification. Additionally, a multi-scale module is introduced to enhance detection performance. Figure \ref{vidSteg} illustrates the the overall framework of PUSN. The experimental results demonstrate that, in comparison to existing methods, the proposed approach effectively detects multiple \ac{PU}-based steganography methods and achieves superior detection accuracy across various embedding rates, with an accuracy of up to 99.02\%. This accuracy is 3.07 percentage points higher than the method it is compared to.}

 {Challenges are found in achieving desirable outcomes when dealing with \ac{VSOFD} because of the following constraints: (i) the absence of specialized features tailored for effective \ac{VSOFD} processing, and (ii) the absence of a dedicated deep network architecture explicitly designed for \ac{VSOFD}. To tackle the latter constraints, the work in \cite{gan2023video} employed a residual-based steganalysis feature (RSF) to extract \ac{VSOFD}-specific features from the spatial-temporal-frequent domain. The RSF feature is then used to form the residual-based steganography feature vector group (RSFVG), which serves as the input of the \ac{DL} network, called parallel-DenseNet-concatenated-LSTM (PDCL). By extracting key clues of video frames from RSF, instead of raw frame images, the proposed approach can more effectively learn and identify forgery frames. Therefore, steganalysis plays a crucial role in detecting object forgery in video surveillance. The framework attains an exceptional F1 score of 90.33\%, representing a substantial enhancement of nearly 8 percentage points compared to existing \ac{SOTA} methods.}

\subsection{Audio/Speech steganalysis}

Due to significant progress in DL methods, audio steganalysis has recently utilized these models and typically outperforms conventional techniques. This section introduces some recent studies on audio steganalysis using DL models. Table
\ref{tab:speechText} summarizes many of the detailed \ac{DL}-based steganalysis schemes for speech, text, and other data types.

\subsubsection{RNN/LSTM-based {methods}}

The RNN architecture is customized to identify and extract valuable insights from sequential or time-series data. One such application is analyzing audio files, such as speech recordings, to identify patterns of correlation. To address the need for high detection accuracy and quick response time in detecting hidden messages in \ac{VoIP} streams, Lin et al. \cite{lin2018rnn} proposed an RNN model. Qiu et al. \cite{qiu2022steganalysis} tackle the issue of detecting \ac{QIM}-based steganography in short or low-embedding-rate speech streams. To accomplish this, they propose a steganalysis model that uses distributed representations. The model has a code-word embedding layer that captures distributed representations in a denser space. It also includes a bidirectional Bi-LSTM layer and a gated attention mechanism to provide contextual distribution features with superior generalization capabilities. Lastly, they designed a \ac{MLP} classifier to differentiate between cover and steganographic objects. The results of their experiments indicate that their model can successfully detect \ac{QIM}-based steganography in \ac{AMR} speech streams and surpasses previous \ac{SOTA} models. Wang et al. \cite{wang2022steganalysis} propose a steganalysis technique capable of identifying various types of steganography concurrently in compressed speech. Their approach involves a code-word-distributed embedding module that condenses compressed code words into a condensed feature representation. The technique then employs two correlation mining modules: a global-guided module (comprising Bi-LSTM and multi-head self-attention) and a local-guided module (consisting of convolution blocks and \ac{CBAM}) to extract correlation changes before and after steganography at both the global and local levels. The detection outcome is then determined using fully connected layers.

\subsubsection{CNN/DBN-based {methods}}

Chen et al. introduce the first audio steganalysis method based on \ac{CNN} in 2017 \cite{chen2017audio}. They propose a \ac{CNN} architecture that operates in the time domain to identify audio steganography. The layers of the network are structured to eliminate audio content and flexibly detect the subtle alterations made by $\pm 1$ LSB-based steganography. In addition, they apply a combination of convolutional layers and max pooling to perform sub-sampling, ensuring effective abstraction and avoiding over-fitting. Similarly, Lin et al. \cite{lin2019audio} also introduced a CNN-based audio steganalysis in the time domain. They use a \ac{HPF} layer to extract the residual signal and generated hierarchical representations of the input through six sets of layers. The results show the potential and effectiveness of the proposed method in detecting LSB matching and \ac{STC} steganographic algorithms at various embedding rates.
Wang  et al. \cite{wang2018cnn} introduced a new method for detecting hidden data in \ac{MP3} files using a CNN-based approach. This technique shows excellent results when applied to the  \ac{EECS} steganography algorithm. The authors in \cite{duwinanto2019steganographic} aimed also to develop a CNN-based \ac{MP3} steganalysis technique. 
They focus on identifying messages that were embedded using the MP3Stego and \ac{EECS} algorithms. Their approach involves classifying the files based on the algorithms used and the approximate length of the message.  To accomplish this, they combine the \ac{QMDCT} audio feature with \ac{CNN} architecture.

Yang et al. \cite{yang2020hierarchical} developed a Hierarchical Representation Network for steganalysis of \ac{QIM} steganography in low-bitrate speech signals. Their model uses a \ac{CNN} and three levels of attention mechanisms to selectively focus on important content in speech frames. Results show their method outperforms existing approaches in steganalysis, particularly in detecting short and low-embedded speech samples, and is more time-efficient for real-time online services.

 { Paulin et al. \cite{paulin2016audio} constructed a \ac{DBN} by arranging multiple \acp{RBM} in a stacked manner where the initial visible layer takes in the input data, and the next \ac{RBM} is trained with the preceding one's hidden layer. They use \acp{MFCC} as features and a \ac{DBN} as a classifier to address two distinct objectives. The first task is to detect whether a signal had a concealed message, while the second task is to determine the steganographic algorithm used to embed the data. The authors compare their \ac{DBN} architecture to \ac{SVM} and \ac{GMM} and show that, in general, \ac{SVM} lead to the best outcomes for the first task, while \ac{DBN} outperforms \ac{SVM} and \ac{GMM} in terms of accuracy in the second task.}  Later, an improvement of this technique using \acp{EA} was presented in \cite{paulin2016speech}. The \acp{EA} are exploited to train the \ac{RBM} used to construct the \acp{DBN}.

\subsubsection{Residual neural networks-based {methods}}

\begin{figure}[h!]
    \centering
    \includegraphics[scale=1.2]{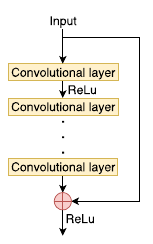}
    \caption{ The diagram of a residual block, the shortcut is depicted by the arch encircling the layers.}
    \label{resBlock}
\end{figure}

\Acp{ResNet} are specifically created to address the vanishing gradient problem that arises while training neural networks that are very deep. They solve this problem by incorporating skip connections or shortcuts between layers. These shortcuts modify the function $F (x)$ that maps data by adding the identity function x. Consequently, the mapping function becomes $F (x) + x$, which is simpler to optimize as it is related to the input data. Figure \ref{resBlock} shows a diagram of a residual block \cite{tabares2020digital}.

\begin{figure}[h!]
    \centering
    \includegraphics[scale=1.2]{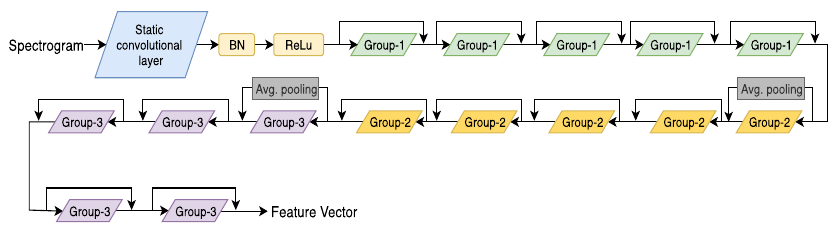}
    \caption{ The S-ResNet design for speech steganalysis: It comprises of 31 convolutional layers and a non-trainable layer, divided into three groups, each containing ten layers. The initial group (Conv-1) utilizes ten filters per layer, the second group (Conv-2) uses 20 filters, and the third group (Conv-3) applies 40 filters. The layers in the first group have a stride of 1 in both dimensions, while those in the second and third groups have a stride of 2 in the vertical and horizontal axes, respectively \cite{tabares2020digital}.}
    \label{resnet}
\end{figure}

In  {Ren et al.} \cite{ren2019spec}, a steganalysis scheme was designed using a ResNet architecture called S-ResNet. The authors use the spectrogram of an audio signal as input to the neural network. The spectrogram is a visual representation of the frequency components of a signal over time, and it can be seen as an image of size $n \times m$, where $n$ is half the frame size, and $m$ is the number of frames for a given window size. The S-ResNet architecture, shown in Figure \ref{resnet}, is composed of 31 convolutional layers, and the first layer has four fixed filters, while the other filters in the network are updated during training. All filters have a size of 3 $\times$ 3, and after each convolutional layer, there is \ac{BN} and a \ac{ReLU} layer. The skip connection defines the residual blocks and allows the training a \ac{DNN} and two average pooling layers reduce the data volume after every five residual blocks. There is a global average pooling layer at the end of the network to flatten the volume onto a 1-D vector of length 40, called the feature vector. The first four filters of the network are handcrafted to amplify the noise of the signal, which can be associated with hidden messages. Using this scheme, the authors achieve high levels of accuracy for both \ac{AAC} and \ac{MP3} audio formats. The testing is conducted using various window sizes to generate the spectrogram. The feature vector is then passed to an \ac{SVM} to execute the final training and classification steps.  Overall, this work demonstrates the benefit of using a \Ac{ResNet} architecture to extract complex and discriminative features from audio data for steganalysis purposes.

\subsection{Text steganalysis}

Generally, linguistic steganalysis methods involve the extraction of semantic features from texts, followed by an analysis of the differences in these features between cover and stego texts to determine if a sentence contains hidden information. As a result, some advanced linguistic steganalysis techniques have been proposed, which have yielded remarkable results. For instance, reference \cite{xu2021small} used pre-trained \ac{BERT} language mode to act as an embedder and generate initial word representations. It was then recognized that any local semantic changes before or after steganography could alter the overall semantics of a sentence. Consequently, a feature interaction module was developed to model the mutual effects between the extracted local and global semantic features.

\subsubsection{DNN/CNN-based {methods}}

According to Wen et al \cite{wen2019convolutional}, a CNN-based model can be used for text steganalysis, which captures intricate dependencies and automatically learns text feature representations. To improve performance, a decision strategy is also suggested for detecting long texts. Firstly, the word embedding layer extracts both the semantic and syntactic features of words. Secondly, different-sized rectangular convolution kernels are used to learn sentence features. This method is effective not only for exploring various text steganography algorithms but also for analyzing texts of different lengths. On the other hand, Xiang et al. \cite{xiang2020convolutional} proposes a two-stage CNN-based method for text steganalysis. The first stage is a sentence-level \ac{CNN} that consists of a convolutional layer with multiple convolutional kernels of different window sizes, a pooling layer, a fully connected layer with Dropout, and a soft-max output. In this way, the layer can handle variable-length sentences and obtain two steganographic features per sentence. The second stage is a text-level \ac{CNN} that utilizes the output of the first stage to determine whether the detected text is steganographic or not. The average accuracy of this method is 82.24\%. Yang et al. \cite{yang2018ts} proposed a semantic-based text steganalysis technique called TS-CNN. This method employs \ac{CNN} to extract high-level semantic features of texts and detects subtle variations in the semantic space before and after embedding hidden information to accurately identify steganographic texts. Experimental outcomes indicate that the proposed model attains almost 100\% precision and recall.

 {Yuanfeng, et al.} \cite{luo2021creative} employs an optimized \ac{DNN} to uncover the connection between hidden business information and structured data, taking into account the effects of data generalization and hidden data optimization. The idea is based on two models for feature fusion and classification, referred to as \ac{DNNITS} model, which involves concatenating each line of text message input, combining the entire message into a single sentence, and then utilizing the previously described model to conduct feature extraction.


\subsubsection{RNN/LSTM-based {methods}}

The embedding of hidden information causes distortion of the conditional probability distribution in stego text that is automatically generated. A text steganalysis algorithm presented in paper \cite{yang2019ts} suggests using \ac{RNN} to identify these differences in feature distribution and classify them into cover text and steganographic text. The model's experimental results indicate that it has high accuracy in detecting steganographic text and can estimate the amount of information embedded in the generated stego text using subtle differences in text feature distributions.  In \cite{li2021text}, a text steganalysis approach is proposed that aims to enhance low-level features in the feature vector and better associate them with steganographic information in the generated text. The approach includes two parts: dense connectivity and feature pyramid. The method is based on densely connected \ac{LSTM} networks with a feature pyramid. First, the words in the text are mapped to a semantic space with hidden representations to better utilize semantic features. Then, stacked bidirectional \ac{LSTM} networks are used to extract semantic features at different levels. Finally, the semantic features at all levels are fused, and a sigmoid layer is used to determine whether the text is steganographic or not. To capture the long-term semantic information of texts, the Bi-LSTM architecture is employed in \cite{niu2019hybrid}. Asymmetric convolution kernels with varying sizes are also applied to extract the local relationships between words. Yi et al. \cite{yi2022exploiting} proposed two methods to improve the performance of LSTM-based text steganalysis models. The first method involves pre-training a language model based on \acp{LSTM}, while the second method involves pre-training a sequence autoencoder. The parameters learned during pre-training are used to initialize the corresponding LSTM-based steganalysis model. The study shows that both methods result in improved performance compared to randomly initialized \acp{LSTM}, and the convergence speed is significantly faster.

\subsubsection{GNN-based {methods} }
Recently, there has been growing attention towards applying \ac{DL} techniques to graph data, leading to the popularity of \ac{GNN}. \acp{GNN} are \ac{DL}-based methods that operate on the graph domain. These networks are designed to learn representations of graph data and can effectively handle both node-focused and graph-focused tasks. Unlike \acp{CNN}, \acp{GNN} captures the relationships within graphs by passing messages between nodes and maintaining a state that encodes information from any accessible neighborhood through edges \cite{zhou2020graph}. Additionally, \acp{GNN} possess an enhanced ability to use global information. Due to their convincing performance, \acp{GNN} have been recently introduced for steganalysis in several works \cite{wu2021linguistic,fu2022hga,liu2022graph}. 

In  \cite{wu2021linguistic}, a GNN-based approach for text steganalysis was proposed. This approach differs from traditional methods that treat texts as sequences, as it models texts as graphs with associated information. For each text, a directed weighted graph is created, which is then used to train a graph convolutional network for feature extraction and text classification. To enhance the use of global information, the model adopts a globally-shared matrix that captures the correlation strengths between words. In \cite{fu2022hga}, the graph channel attention module is used to extract the most notable distinguished text representation feature on the dimension level of a gated \ac{GNN}'s node.

\begin{figure}[h!]
    \centering
    \includegraphics[scale=1.1]{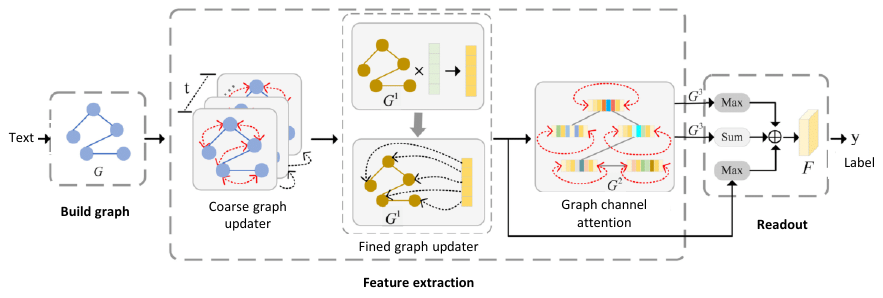}
    \caption{ An example of a scheme that employs \ac{GNN} and attention layer \cite{fu2022hga}.}
    \label{hga}
\end{figure}

According to Figure \ref{hga}, the Gated \ac{GNN} is uses as the primary means to update the node representations of the entire graph with contextual information at the graph level. Building on this coarse graph updater, a graph attention mechanism is introduced as a secondary means to extract detailed features and obtain the word-level representation. The graph channel attention module is then employed to identify the most significant dimension in the node embedding. Ultimately, the readout function is devised to merge the features from various levels and classify the text representation as either cover or stego.

\begin{table}[h]
\caption{Cover-based steganalysis includes audio, text, and other cover types. The type of cover could be identified from the data set used. When compared to much \ac{SOTA} work, only the two best methods are mentioned. }
\label{tab:speechText}
\scriptsize
\begin{tabular}{lm{1.3cm}m{2.5cm}m{1.5cm}m{1.5cm}m{1.5cm}m{6cm}}
\hline
 Paper& DL method  & Steganographic algorithm & Compared to & Data set &  {Metric used} & Observations\\
 \hline

\cite{lee2020deep} & CNN   & LSB, Hide4PGP, Steghide  & \cite{boroumand2018deep,lin2019audio} &TIMIT  &  {AUC, \newline 
 accuracy}  &   SRNet, and BSNet architectures are employed.  {BSNet is the best, achieving 89.14\% accuracy and 93.68\% AUC.} \\[0.3cm]

\cite{paulin2016audio} & DBN  & StegHide, Hide4PGP, FreqSteg & SVM, GNN &Noizeus &  {Accuracy} & Achieved higher accuracy (5\% more) than SVM and GMM in identifying the steganographic method used. \\[0.3cm]

\cite{lin2018rnn}  & RNN & CNV-QIM & IDC, SS-QCCN  & Created &  {Accuracy} & Achieved a detection accuracy of 90\% even for a 0.1-second sample and a testing time that averaged 0.15\% of the sample length. \\[0.4cm]

\cite{chen2017audio}  &  CNN  & LSB & Hand-crafted features &Wall Street Journal &  {Accuracy} & Achieved
an 88.3\% classification accuracy for signals with a 0.5 bps embedding rate. \\[0.4cm]

\cite{lin2019audio} & CNN  &  LSB
matching, STC  & \cite{chen2017audio} & TIMIT &  {Accuracy} & \ac{HPF} is used.  {The final output consisted of 215 features, which were then input into a binary classifier, resulting in an accuracy of 90.50\%.} \\[0.4cm]

\cite{wang2018cnn} & CNN   & HCM-Gao, HCM-Yan, EECS &  Hand-crafted features& Created &  {Accuracy} & QMDCT coefficients matrix is selected as the input data. Accuracy can reach up to 95.35\%. \\[0.3cm]

\cite{yang2020hierarchical} & CNN  & QIM & \cite{lin2018rnn,yang2022real}& Created in \cite{lin2018rnn} &  {Accuracy} & The detection accuracy reached 87.14\% of various models. Use nearly a quarter of the parameters used by the CSW method. \\[0.3cm]

\cite{ren2019spec} & ResNet  & LSB-EE, MIN, SIGN, MP3Stego &  \cite{wang2018cnn,jin2017steganalysis} &Created &   {Accuracy} & Accuracy reaches  94.98\% for \ac{AAC} format, and 99.93\% on
the \ac{MP3} format. \\[0.3cm]

\cite{xu2021small}& CNN &  Tian-Fang, GLMA  & \cite{wen2019convolutional,yang2019ts} & IMDB Movie Review, News, Twitter &  {Accuracy} & Bidirectional \ac{GRU} is used to extract global features from both directions.  {The detection depends on the specific scenario, and it can reach up to 97.4\%.}  \\ [0.3cm]

\cite{yang2019ts}& RNN &  Luo, Fang &  Statistical methods, Genetic algo.   &T-Steg &  {Precision, recall, accuracy} & In some situations, it can achieve around 100\% accuracy, precision, and recall. \\[0.3cm]

\cite{li2021text}&Capsnet &  RNN and variable-length coding &\cite{yang2019fast,yang2019ts} & IMDB Movie Review, News, Twitter &  {Accuracy} & Achieves 92\% accuracy at 1-3 bit/word, 7\% better than other neural networks. At 4-5 bit/word, the detection accuracy was over 94\%.\\[0.3cm]

\cite{niu2019hybrid}& BILSTM-CNN & T-Lex, Tina-Fang & \cite{wen2019convolutional,yang2019ts} &Twitter, IMDB Movie Review, Gutenberg &  {Accuracy, Recall, Precision.} & A method called R-BILSTM-C, can achieve an accuracy of 100\% in variable-length sentences. \\[0.3cm]

\cite{yi2022exploiting}  & LSTM  & RNN-Steg, Bins & FCN, CNN, GNN, RNN & IMDB Movie Review, Twitter &  {Accuracy} & Achieves a max accuracy of 91.75\% for MOVIE data set and 96.43\% for Twitter data set.\\[0.3cm]

\cite{fu2022hga}& GNN & RNN, VAE  & \cite{yang2020ts,xu2021small} & IMDB Movie Review, Twitter &  {Accuracy, F1-score} & It is a linguistic steganalysis method, called HGA.  { Detection can achieve an accuracy of 93.5\% and an F1-score of 92.3\%. }  \\[0.3cm]

\cite{qiu2023separable}& CNN  & Geiser, Miao &\cite{ren2015amr,tian2019steganalysis} & Created in \cite{lin2018rnn} &  {Accuracy} & The method called SepSteNet, is based on separable CNN. The dual-stream pyramid module enhances  {accuracy by up to 2.98\%} by mitigating the negative impact of sample content.  \\

\cite{wu2022amr}& Adversial Bi-GRU &  Huang \cite{huang2012steganography} & \cite{ren2016amr}   & Created &  {Accuracy} & For normal and adversarial samples, it achieves accuracy of 96.73\% and 95.6\%, respectively. \\

\cite{bae2018dna} &RNN&Developed in \cite{mitras2013proposed}&Adaptive boosting, SVM,   Random forest & Human UCSC-hg38&  {Accuracy} &Developed for unusual cover type (\ac{DNA}).  {The detection rate reaches 99.93\% for both sides (Intron and Exon regions). Exceeds the BLAST method by 15.93\%.} \\

\cite{bae2020dna}&RNN+CNN&Developed in \cite{mitras2013proposed}&SVM, Adaptive boosting and Random forest classifiers &Human UCSC-hg38&  {Accuracy} & Used \ac{DNA} cover type.  {The model successfully identified all alteration rates in all DNA regions when tested with a sample length of 18,000. The accuracy difference between perturbed and non-perturbed sequences in the intron region is around 0.1.}\\[0.4cm]

\cite{zhao2020bns} &CNN& Methods \cite{miller2008steganography,lucena2006covert,rowland1997covert} & LeNet, AlexNet  & Created &  {Accuracy} & Developed for unusual cover type (IPv6 network).   {Exceptionally accurate at 99.98\% while maintaining a low level of time complexity.}\\[0.2cm]

\hline
\end{tabular}
\end{table}

\subsection{Other cover-based steganalysis}

Zhao and Wang \cite{zhao2020bns} use \ac{CNN} to identify network steganographic techniques used in the IPv6 network. They propose a unified steganalysis model based on \ac{CNN}, named BNS-CNN, to detect multiple network storage steganographic algorithms. The model preprocesses network traffic and divides it by field to maintain the feature's integrity to the maximum extent, resulting in a matrix. The feature extraction is achieved through multiple convolution kernels and K-max pooling to hasten model convergence, while the fully connected layer is designed to improve feature integration and enhance the model's robustness. Research findings demonstrate that the BNS-CNN model achieves 99.98\% detection accuracy with low time complexity and favorable generalization performance. In the field of steganalysis, they have been utilized for analyzing hidden information in \ac{DNA} sequences  \cite{bae2018dna, bae2020dna}. Bae et al. \cite{bae2018dna,bae2020dna}  proposed using \ac{RNN}  to detect malicious \ac{DNA} sequences. The concept involves utilizing \ac{RNN}  to learn intrinsic distributions and detect variation in distributions using a classification score to predict whether a sequence is coding or non-coding. The experimental results show that this approach provides better detection performance compared to common detection approaches, namely frequency analysis-based methods.

\section{{Advanced} technique-based deep steganalysis}
\label{sec5}
In this section, steganalysis schemes are categorized according to their utilized \ac{DL} technique, rather than the type of cover. { This includes methods that use advanced \ac{DL} techniques such as \ac{DTL}, \ac{DRL}, and hybrid \ac{DL} techniques.}

\begin{table}[]
\caption{{Advanced} technique-based deep steganalysis.  { {In DTL-based steganalysis methods, any improvements achieved are in comparison to full training}}. When compared to much \ac{SOTA} work, only the two best methods are mentioned. }
\scriptsize
\label{tab:techType}
\begin{tabular}{m{0.5cm}m{1.2cm}m{2cm}m{1.8cm}m{1.5cm}m{1cm}m{7.5cm}}
\hline
 Paper& DL Type  & Steganographic algorithm & Compared to & Data set &  {Metric used}& Observations\\
 \hline
 \cite{zhang2022dataset} & DTL (SDA)& S-UNIWARD, WOW, HUGO& \cite{boroumand2018deepieee,zhang2020cover} &BOSSbase, UCID, MIRFlickr25K, DIV2K &  {Accuracy} & Accuracy boosted by 5.4\%, 8.5\%, and 8.0\% at an \acs{ER} 
 equal to 0.4, 0.2, and 0.1 \ac{bpp}, respectively, in the presence of data set mismatch.\\[0.4cm]

\cite{padmasiri2021impact}& DTL  & J-UNIWARD, JMiPOD, UERD & \cite{boroumand2018deepieee,zhang2019new} & ALASKA  \#2 &  {F1-score} & \ac{DTL} employed the latest and triumphant ImageNet classification models such as EfficientNet, MobileNetV3, DenseNet, and MixNet.  {EfficientNet3 (3 classes) can reach an F1-score of 84\%.}  \\ [0.3cm]

 \cite{yousfi2020imagenet} & DTL  & J-UNIWARD, JMiPOD, UERD & \cite{boroumand2018deepieee} & ALASKA  \#2 &  { Weighted AUC} & EfficientNet, MixNet, and ResNet pre-trained on ImageNet and adapted for JPEG steganalysis offered better performance,  {achieving a weighted AUC (wAUC) of 93.85\%, compared to the SRNet model trained from scratch, which achieved a wAUC of 92.27\%.} \\
 
 \cite{ahmed2022image} & DTL  &  J-UNIWARD, nsF5, UERD & AlexNet,
SRNeT &IStego100K &  {Accuracy} & This approach exploit pretrained AlexNet and achieves an accuracy of 99\%. \\

 \cite{zhan2017image}  & DTL  & BOW & Without DTL &BOSSbase&  {Precision} & This method pertains to the field of image forensics. It has the ability to identify five distinct types of image alterations with an average precision of 97.36\%. \\[0.4cm]
 
 \cite{dengpan2019faster} & DTL & WOW, \newline S-UNIWARD & \cite{fridrich2012rich,ye2017deep} &BOSSbase &  {DER} & The suggested technique demonstrates improved effectiveness in comparison to \cite{fridrich2012rich, ye2017deep} for certain types of payloads,  {achieving a low \ac{DER} of around 45.30\%, which is better to both parameter multiplexing, and fine-tuning, where the \ac{DER} is around 47.31\%.} \\[0.4cm]
 
\cite{el2018improved} & DTL  & WOW, \newline S-UNIWARD, HUGO & Without DTL & BOSSbase &  {Accuracy} & Exhibits faster convergence for low payloads and higher detection accuracy for high payloads,  {with a classification rate that can reach up to 96\%, showing a 70\% improvement when DTL is not adopted.} \\

\cite{peng2021real} & DTL &   Block-based, \newline VLC-based  & \cite{niu2019hybrid,yang2020linguistic} &Twitter, IMDB Movie Review &  {Accuracy} & Achieves superior performance compared to other tested DNN-based methods, with a maximum accuracy of 97.20\%, regardless of the embedding capacity and steganographic algorithm used in both data sets. \\[0.4cm]

\cite{mustafa2019accuracy} & DTL  & S-UNIWARD, WOW & \cite{fridrich2012rich,kodovsky2011ensemble} &BOSSbase &  {Accuracy} & Surpasses the approach being compared to it, as it displays an average metric improvement of 7.4\%, 5.3\%, 4.1\%, and 2.8\%. \\[0.4cm]

\cite{yang2020reinforcement}& DRL  & J-UNIWARD, UERD & \cite{boroumand2018deepieee,xu2017deep}  & BOSSbase, BOWS2  &  {Accuracy} &  Outperform \ac{SOTA} methods  { by achieving an improvement in detection accuracy ranging from a minimum of 1.5\% to a maximum of 16\%.} \\

\cite{hu2019digital}& DRL & J-UNIWARD & \cite{xu2017deep,denemark2016steganalysis}&BOSSbase&  {Accuracy} & Achieves a classification with the lowest error rate of 0.0807 after the fifth iteration.  {For J-UNIWARD, it outperforms Xu's method by 0.5\% to 1\% in detection accuracy.}\\

\cite{ni2019selective}&DRL& S-UNIWARD, \newline J-UNIWARD & PEP \cite{qian2015pareto} & BOSSbase&  {Accuracy} & It can improve the performance of GFR, SRM, and maxSRMd2 at varying embedding payloads.  {The performance exceeds that of PEP by approximately 0.57\% while utilizing the same features.}\\

\cite{cohen2020assaf}& DAE+SNN  & J-UNIWARD & \cite{boroumand2018deepieee,xu2017deep}& BOSSbase &  {Accuracy} & Achieves detection accuracy  between 80.3\%  and 99.3\%. \\

\cite{kang2020classification}&  CNN+DRN  & WOW, PVD, LSB, \newline S-UNIWARD  & Single CNN &BOSSbase &  {Accuracy} &  Outperforms the single CNN approach. Accuracy increased by 17.71\% \\

\cite{gong2019recurrent}& Bi-LSTM+CNN  & \Ac{FCB} & \cite{miao2014steganalysis,ren2015amr} & Created in \cite{lin2018rnn} &  {DR} & Correct \ac{DR} increases by 10\% with 100 ms sample at 20\% embedding rate. \\

\cite{yang2019steganalysis} & CNN+LSTM &CNV-QIM  & \cite{lin2018rnn}&Created in \cite{lin2018rnn}&  {Accuracy} & Improves detection accuracy at low embedding rates and achieves an accuracy above 95\% when the embedding rate is over 40\%.\\

\cite{hu2021detection} & DTL &CNV-QIM, PMS, CNV-QIM+ PMS &\cite{lin2018rnn}&Created in \cite{lin2018rnn}&  {Accuracy} &Achieves satisfactory accuracy in detecting CNV-QIM and performs the best among the compared methods in detecting PMS and CNV-QIM+PMS.  {After fine-tuning, the accuracy improved by 3.7\% for English speeches at an embedding rate of 30\%, compared to the accuracy without fine-tuning.}\\

\hline
\end{tabular}
\end{table}

\subsection{Deep transfer learning-based {methods} }

\Ac{DTL} is a method of improving generalization in a learning task by making use of knowledge gained from another, previously learned task \cite{himeur2023video,kheddar2023deep,kheddarASR2023}. In traditional \ac{ML}, separate data sets and training processes are used for different tasks (source task $\mathbb{T}_{S}$ and target task $\mathbb{T}_{T}$), and no knowledge is transferred or accumulated between models \cite{sayed2022deep}. TL, on the other hand, makes use of knowledge such as features and weights gained from previously trained models in a source domain $\mathbb{D}_{S}$ to train new models in a target domain $\mathbb{D}_{T}$. Additionally, \ac{DTL} can be used even when there is less data or no label information available in the target domain. \textit{Inductive} \ac{DTL} refers to transfer knowledge learned from a $\mathbb{D}_{S}$ to a $\mathbb{D}_{T}$ that are not always equal, 
i.e. tasks and data sets are different. \textit{Transductive} \ac{DTL} refers to transfer knowledge learned from a $\mathbb{D}_{S}$ to a $\mathbb{D}_{T}$ that is totally different, and tasks are the same. Training \ac{DL} models is a challenging task that requires significant processing time and data. As a result, pre-trained models have been used as feature extractors to reduce training time. Besides that, \ac{DTL} is used to solve the problem of data and/or label scarcity, and computation cost \cite{sohail4348272deep}.

In steganalysis, \ac{DTL} plays a vital role in feature extraction and model adaptation. It allows transferring knowledge learned from one domain (e.g., image classification) to another domain (e.g., image steganalysis). Pre-trained models serve as effective feature extractors, and \ac{DTL} addresses issues related to data and label scarcity. \ac{DTL} is particularly valuable when dealing with data set mismatches.

For example,  {Taha et al.} in \cite{ahmed2022image} introduced an image steganalysis technique that applies the AlexNet \ac{CNN} model. The method consists of three steps: (i) data collection and preparation, (ii)  extraction of distinctive features using the AlexNet model, and (iii) training a random forest classifier with the feature vectors to detect binary classifications of cover and stego.
Reference \cite{yang2017jpeg} use a pre-trained model, DenseNet, along with a conventional method to decrease detection error rates, resulting in ensemble approaches within the domain context. The research in \cite{padmasiri2021impact} analyze approaches that are not commonly explored in the image steganalysis field. Specifically, it focused on \ac{DTL} using recent ImageNet models such as EfficientNet, MobileNetV3, DenseNet, and MixNet, and try to develop ensemble models using these pre-trained architectures. Similarly, \cite{yousfi2020imagenet} explores the use of pre-trained deep architectures in computer vision, including EfficientNet, MixNet, and ResNet, for steganalysis. These models, which have been pre-trained on ImageNet, can be easily adapted for JPEG steganalysis and offer better performance than \acp{CNN} that are specifically designed for steganalysis, such as the SRNet that is trained from scratch. The work of  Zhang et al. \cite{zhang2022dataset} introduces a technique for addressing the problem of data set mismatch in \ac{DL}-based steganalysis. This is achieved through a feature-guided \ac{SDA} framework (Transductive \ac{DTL}), using the SRNet network as a baseline. To start with, both the source and target domains are split into subdomains based on class, and the relevant subdomains are aligned using \ac{SDA} to address any distributional discrepancies. A guiding feature is then created to improve the accuracy and precision of the subdomain division. Ozcan and Mustacoglu \cite{ozcan2018transfer} propose using \ac{DTL} to perform image steganalysis in the spatial domain with a\ac{DRN}. They train a ResNet50 model using the Keras \ac{DL} framework on the ImageNet data set. Detailed experiments were conducted and showed improved results in detecting stego images across various payload data sets. However, the scheme has not been compared to any other steganalysis detectors,  and it has not been evaluated using a standard data set. Two \ac{DTL} are used in \cite{dengpan2019faster}, namely parameter multiplexing and fine-tuning, to enhance the overall efficiency and reduce computation cost and time. Rabii El Beji et al. \cite{el2018improved} propose a new technique that combines kernels catalyst and \ac{DTL} to guarantee quick convergence for small payloads through weight propagation. The results and findings of some of \ac{SOTA} steganalysis schemes, which are based on \ac{DTL}, are detailed in Table \ref{tab:techType}.





\subsection{Deep reinforcement learning-based {methods} }
Reinforcement learning (RL) is a type of \ac{ML} where an agent learns to make decisions in an environment by receiving feedback in the form of rewards or punishments. The main goal is to maximize the total reward accumulated over time by choosing the best action based on the current state of the environment. \ac{DRL} is a variation of RL that uses \acp{DNN} to represent the agent's decision-making process. By using \acp{DNN}, the agent can recognize patterns and make more complex decisions, leading to better performance in complex environments.

\begin{figure}[t]
   \centering
    \includegraphics[scale=0.75]{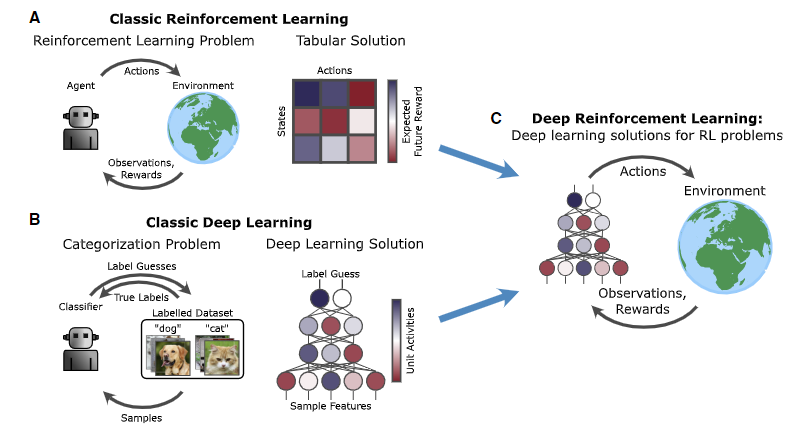}
    \caption{RL, DL and DRL \cite{botvinick2020deep}. }
    \label{fig:DRL}
\end{figure}

Figure \ref{fig:DRL} provides a clear illustration of the difference between RL, \ac{DL}, and \ac{DRL} \cite{botvinick2020deep}. In subfigure (A), the RL problem is shown, where the agent selects actions, receives observations and rewards from the environment, and tries to maximize the long-term reward. A tabular solution is also shown, where the agent learns the expected long-term reward associated with taking each action in each state. In subfigure (B), the supervised learning problem is depicted, where the agent receives unlabeled data samples and tries to guess the correct labels. A \ac{DL} solution is also shown, where the features of a sample are passed through multiple layers of artificial neurons to predict the correct label. In subfigure (C), \ac{DRL} is shown, where a neural network is used as an agent to solve an RL problem. This method has been found to perform well in complex environments by learning appropriate internal representations that generalize well to new states and actions.

\Ac{DRL} enhances steganalysis by training agents to autonomously scrutinize digital media, taking actions guided by environmental feedback based on detection accuracy. \Ac{DRL}'s impressive ability to discern intricate patterns and relationships within media content adapt to diverse steganographic techniques, and continuously learn and improve, positions it as a vital asset for uncovering concealed information in complex and dynamic digital landscapes, transcending traditional steganalysis approaches.


Various DTL-based steganalysis methods have been proposed in the existing literature \cite{hu2019digital, ni2019selective, yang2020reinforcement}.  Hu et al. \cite{hu2019digital} developed a steganalysis technique using \ac{DRL} and visual attention to detect adaptive steganography in JPEG images. The method selects a region of interest using visual attention, then uses \ac{DRL} to create a summary region, guiding the \ac{DL} model to focus on favorable regions. The detection ability is enhanced by replacing misclassified training images with their corresponding summary regions. Ni et al. \cite{ni2019selective} introduced a deep Q-network-based ensemble technique for image steganalysis that utilizes \ac{DRL} and \ac{CNN} to enhance the model's generalization performance and decrease the ensemble's size. Their experimental findings indicate that this approach is effective for optimizing the ensemble classification of image steganalysis in both spatial and frequency domains.

\subsection{Hybrid steganalysis methods}

In the pursuit of advancing steganalysis accuracy, researchers have explored hybrid approaches that combine diverse \ac{DL} models \cite{kang2020classification, gong2019recurrent, yang2019steganalysis, hu2021detection, peng2021real}. These innovative strategies are engineered to harness the distinct strengths of multiple models, resulting in a synergistic enhancement of the effectiveness of hidden message detection. The fundamental rationale behind hybrid steganalysis methods is to capitalize on the unique capabilities offered by different DL architectures and frameworks. Each DL model may excel in certain aspects of feature extraction, pattern recognition, or domain-specific knowledge, enabling steganalysis to benefit from a holistic analysis approach that considers a broader spectrum of features and characteristics within the data. This approach mitigates the limitations of relying solely on a single DL model, which may be specialized for specific tasks but may struggle with other types of steganography. Through integration, hybrid models can adapt to varying steganographic techniques and cover media, making them more versatile and robust.


For instance, Kang et al. \cite{kang2020classification} proposed an image steganalysis approach that can categorize images into five classes, which involves classifying between stego and cover images and identifying four spatial steganographic algorithms.  This approach utilizes a combination of a hierarchical structure of CNN and DRN. Results from experiments indicate that the proposed technique outperforms the single CNN approach. Gong and Yi \cite{gong2019recurrent} suggest using a combination of Bi-LSTM and \ac{CNN} for \ac{AMR} audio steganalysis in their research paper. They use Bi-LSTM to extract contextual representations of \ac{AMR} fixed codebook (FCB) at a higher level, and \ac{CNN} to merge spatial-temporal features for steganalysis. According to the experiment results, this model has demonstrated an increase in the correct detection rate. Yang et al. \cite{yang2019steganalysis} constructed a  unified model  CNN-LSTM network to detect \ac{QIM}-based steganography in \ac{VoIP} streams. The model employs Bi-LSTM to gather long-term contextual information from the carriers, while \ac{CNN} captures both local and global features and temporal carrier features. Results from experiments demonstrate that this model can achieve superior detection of \ac{QIM}-based steganography in \ac{VoIP} streams.
Hu et al. \cite{hu2021detection} proposed an approach for detecting multiple orthogonal steganographic methods (\ac{QIM} and pitch modulation steganography) in low bit-rate \ac{VoIP} speech streams.  The proposed model combines three neural network structures, namely \ac{CNN}, \ac{LSTM}, and the fully connected network (FCN). The findings indicate that this model outperforms the existing steganalysis methods in detecting HPS and met the need for real-time detection, with a processing time of just 0.34 ms for a 10 ms speech frame.  Peng et al. \cite{peng2021real} developed a text steganalysis neural network based on multi-stage transfer learning. The model comprises three stages: fine-tuned BERT models the difference between normal and steganographic texts, steganography-specific knowledge is distilled to two tiny neural networks (a single-layer Bi-LSTM and a single-layer \ac{CNN}) for fast inference, and different semantic features are integrated in a proposed network using the distilled tiny networks.

Xu et al. \cite{xu2021small} proposed a linguistic steganalysis method for detecting hidden information in small-scale multi-concealed scenarios. The authors use a pre-trained BERT language model to address the lack of domain knowledge resulting from limited data sets to provide an initial word-level representation, which is fine-tuned using training data. Local and global semantic features are extracted using \ac{CNN} and \ac{GRU}, respectively, in parallel to avoid deep model and gradient vanishing problems. A feature interaction module is then designed to explore the relationship between these features before a softmax classifier is used to identify sentences with hidden information. The experiments show that this method effectively detects hidden information and reduces the burden of intercepting large-scale data from network platforms. Reference~\cite{yang2020linguistic} proposes a hybrid text steganalysis method (R-BILSTM-C) by combining the advantages of Bi-LSTM and \ac{CNN}. The method captures long-term semantic information of text using BiLSTM and extracts local relationships between words using asymmetric convolutional kernels of different sizes, leading to a significant increase in detection accuracy. Additionally, the paper visualizes the high-dimensional semantic feature space, demonstrating the approach's effectiveness for different text steganography algorithms. Reference~\cite{niu2019hybrid} proposes an LSTM-CNN model for text steganalysis. First, the words are mapped to a semantic space for a better use of the semantic features of the text. Then, \ac{LSTM} and \ac{CNN} are combined to obtain local contextual information and long-range contextual information in stego text. Additionally, the model employs an attention mechanism to identify important cues in suspicious sentences. The model achieves outstanding results in steganalysis tasks.

\section{Research challenges and future directions}
\label{sec6}

\subsection{Research challenges}
It is true that modern \ac{DL}-based detectors have shown significantly better performance in detecting steganographic signals compared to conventional detectors. However, these detectors face limitations when it comes to training on large images, due to hardware constraints. One of the challenges with steganographic signals is that they are often very subtle and difficult to detect. As a result, resizing or cropping the images prior to classification can compromise the accuracy of the detector. This occurs because such operations can distort the signal and make it even harder to detect. Steganalysis based on \ac{DL} is a potential method for finding secret information in voice, text, and imaging signals. However, there are still a number of research gaps and potential future initiatives in this field of study:

\begin{itemize}
    \item \textbf{Adversarial attacks:} To avoid being discovered by steganalysis algorithms based on \ac{DL}, adversarial assaults might modify or distort the data. Additional research is required to create more reliable algorithms that can recognize adversarial threats.
    \item \textbf{Explainability:} Since \ac{DL}-based models may be complex, it can be challenging to comprehend the features and workings of the algorithms that give them the ability to find hidden data. It is possible to increase the interpretability of the outcomes and contribute to the creation of more transparent and reliable systems by creating explainable models.
    \item \textbf{Lack of training data:} The effectiveness of \ac{DL}-based steganalysis models is greatly influenced by the quantity and quality of the training data. The use and generalizability of these models may be constrained by a lack of extensive steganalysis data sets.
    \item \textbf{Transfer learning:} It refers to the process of transferring knowledge from one domain to another one that has either similarities or dissimilarities. The application of \ac{DL}-based steganalysis models across various kinds of data may be made possible by developing transfer learning approaches.
    \item \textbf{Multimodal steganalysis:} It refers to the process of finding concealed information in several forms of data, such as voice, images, and text. The accuracy of detection may be increased, and false positives can be decreased, by using multimodal steganalysis approaches based on \ac{DL}.
    \item \textbf{\ac{MV}-based video steganalysis:}  Developing more effective convolutional neural networks to minimize embedding distortion in adaptive \ac{MV}-based video steganalysis remains a challenging task.
    \item \textbf{Steganalysis that protects privacy:} The usage of \ac{DL}-based steganalysis might cause privacy issues since sensitive data may be exposed during the analysis. Creating privacy-preserving methods may help to secure private information while still allowing efficient steganalysis.
    \item \textbf{Real-time steganalysis:} It is crucial in a variety of applications, including network security and real-time monitoring. One crucial area of study is the creation of real-time \ac{DL}-based steganalysis methods that can deal with massive amounts of data quickly.
    \item \textbf{Steganalysis in real-world scenarios:} In real-world scenarios, steganalysis faces challenges when the collection of media used for steganography differs significantly from the training set used for the classifier. This discrepancy, known as \ac{CSM}, can occur due to variations in image capture conditions, resolution, compression, filtering, and other factors. Additionally, stego source mismatch can lead to imprecise predictions when embedding parameters or stego-system used vary between the training and testing datasets.

\end{itemize}

In steganalysis, it is crucial to have feature extraction that can detect changes in embedding rates to accurately estimate payload. Additionally, a robust network structure is required since videos with the same embedding rate but different bit rates can have different steganalysis features, which needs to be reduced to address compression. Thus, the quantitative steganalyzer needs to learn steganalysis features that correspond to various embedding rates and have a robust network structure.

By addressing these issues, it may be possible to increase the security and privacy of different applications while also increasing the detection's accuracy and dependability. For example, to overcome these limitations of computation cost, researchers are exploring new techniques such as multi-scale processing, which involves analyzing the image at different scales to detect steganographic signals. Another approach is to use more advanced hardware, such as specialized \acp{GPU}, which can handle larger images more efficiently.

To accomplish the task of enhancing feature representation, classifying images in spatial or frequency domains, processing arbitrary images, and obtaining noise generated by the steganographic process more efficiently, it is necessary to create novel \ac{DL} models and design new computational components. It is important to ensure that this process is as automated as possible.

Regarding real-world scenarios, some authors suggest focusing in actor-based detection problems \cite{lerch2023actor}, in which a collection of users are examined to identify whether they are using or not steganography, for a large variety of embedding algorithms and bit rates.

\subsection{Perspectives}

There has been a significant surge in interest and focus among steganalysis researchers towards the development of DL-based techniques for image steganalysis, surpassing the attention given to other cover types. The realm of image steganalysis has witnessed comprehensive exploration, encompassing various facets such as steganalysis based on image pre-processing, the formulation of universal methods applicable across diverse image types, integration of attention layers, establishment of schemes based on noise, and more. For this reason, we propose extending this diversity beyond CNN-based image steganalysis. Given that speech and text can be transformed into 2D representations and later reverted to 1D, this versatility can be harnessed for these cover types as well. For instance, introducing DL-based preprocessing techniques like the TLU layer to 2D speech representations could pave the way for innovative audio steganalysis methods. This proposition not only broadens the scope but also creates opportunities to apply existing image-based steganalysis approaches, such as transform-based and adversarial-based methods, to the domains of speech and text. Furthermore, employing pretrained image models with 2D representations of speech or text has the potential to result in substantial enhancements. These models, having been trained on millions of images, have generated accurate weights that can be leveraged through DTL-based techniques for audio and text cover types steganalysis.

Recently, generative AI has emerged as a transformative technology that enables machines to generate creative and realistic content, such as text, images, and even audio, mimicking human-like creativity \cite{sohail2023using}. One remarkable example of generative AI is ChatGPT, a language model developed by OpenAI, based on the GPT (Generative Pre-trained Transformer) architecture. ChatGPT can understand and generate human-like text, engaging in conversations with users, answering questions, and even providing creative responses \cite{sohail2023decoding}.

Text, speech, image, and video models can developed by OpenAI \footnote{\url{ https://openai.com/}} can be secured against adversarial attacks by incorporating DL-based steganalysis techniques to detect any hidden perturbations or changes within the input data. Similarly, steganalysis can be used to enhance the security of OpenAI's speech models by detecting any hidden audio signals that may be used to inject malicious code or commands. By incorporating steganalysis into their security protocols, OpenAI can enhance the robustness and security of their systems, and help to prevent potential security breaches or attacks \cite{wei2023generative}. For instance, steganalysis can be used to prevent the ChatGPT model from being fooled by hidden messages that may be embedded within the input text. Adversarial attacks can be carried out by malicious actors who insert subtle changes or perturbations to the input text to deceive the ChatGPT model into generating incorrect or malicious output \cite{qammar2023chatbots}. Steganalysis techniques can be used to detect any such hidden messages and enable the system to filter out any suspicious or adversarial inputs before they are processed by the ChatGPT model. This can help to prevent the model from being fooled by deceptive inputs, and ensure that the generated output is accurate and reliable. In essence, steganalysis can be used as an additional layer of security to detect and filter out any malicious or adversarial inputs, thereby enhancing the robustness and security of the ChatGPT model.

On the other hand, OpenAI can play a significant role in aiding the development of steganalysis schemes for image, text, and speech by leveraging its language generation capabilities and knowledge base. In the case of image steganalysis, for example, ChatGPT can assist researchers by providing insights into various steganographic techniques, common patterns, and potential vulnerabilities in image-hiding methods \cite{yi2022alisa}. It can also help generate annotated image datasets for training steganalysis models and offer feedback on the effectiveness of different detection algorithms. For text steganalysis, ChatGPT can simulate conversations involving steganographic content, thereby helping researchers understand linguistic patterns, semantic inconsistencies, and statistical irregularities that can be indicative of hidden  {information \cite{cao2022generative}}. It can generate steganographic texts and assist in evaluating the performance of detection algorithms. Regarding speech steganalysis, ChatGPT can provide guidance on the techniques used to embed information in audio signals, including frequency domain manipulation and phase coding. It can generate speech samples with hidden content, allowing researchers to develop detection methods and analyze audio features that may expose the presence of steganography \cite{sun2023topic}.

\section{Conclusion}
\label{sec7}
The field of steganalysis is rapidly evolving, spurred by advancements in \ac{DL} algorithms. \ac{DL}-based detectors have shown remarkable performance in detecting steganographic signals, outpacing conventional detectors. However, numerous challenges still hinder the full potential of these detectors, including issues related to hardware constraints, adversarial attacks, interpretability, availability of training data, and privacy concerns, among others.
This review highlights the increasing importance of steganalysis using \ac{DL} in the field of information security. DL-based steganalysis has shown promising results in reducing detection error rates and addressing various issues related to random guessing situations. The survey provides a detailed description of the most famous \ac{DL} models used in steganalysis and summarizes the findings of research for all cover types. The survey has provided a comprehensive overview of the topic by discussing the motivations for the research, presenting the background of \ac{DL} and its concepts, and presenting a well-defined taxonomy. The taxonomy classifies existing DL-based steganalysis proposals based on different aspects, which can help researchers and practitioners to better understand the state-of-the-art in this field. 

In addition, the review highlights the importance of using \ac{DTL} and \ac{DRL} in the steganalysis domain. \ac{DTL} can help to transfer knowledge learned from one domain to another, which can improve the performance of steganalysis models. \ac{DRL}, on the other hand, can be used to train steganalysis models that can adapt to different steganographic techniques and cover types, which can further enhance their detection capabilities. All in all, the review underscores the need for further research in DL-based steganalysis to enhance the security of digital communication and data storage systems.


Moreover, the relevance of steganalysis extends beyond the detection of steganographic signals. Notably, it has significant implications for enhancing the robustness and security of generative chatbots and generative AI models. By detecting hidden perturbations within the input data, steganalysis can safeguard these models against adversarial attacks, thereby improving their reliability.

Despite the aforementioned challenges discussed in this review, steganalysis holds immense potential, especially in light of the ubiquity of digital media. By addressing these issues, it is anticipated that steganalysis will play a critical role in enhancing the security, privacy, and accuracy of digital systems, extending from media to advanced ML models. The future of steganalysis thus promises to be exciting, provided the current research directions bear fruit and are further supported by new insights and technological advancements.

\printcredits

\section*{Declaration of competing interest}

The authors declare that they have no known competing financial interests or personal relationships that could have appeared to influence the work reported in this paper.

\section*{Data availability}
No data was used for the research described in the article.

\section*{Acknowledgement}
The first author acknowledges that the study was partially funded by the Algerian Ministry of Higher Education and Scientific Research (Grant No. PRFU--A25N01UN260120230001).

The fourth author acknowledges the funding obtained by the Detection of fake newS on SocIal MedIa pLAtfoRms (DISSIMILAR) project from the EIG CONCERT-Japan with grant PCI2020-120689-2 (Government of Spain), the PID2021-125962OB-C31 ``SECURING'' project funded by the Ministerio de Ciencia e Innovación, la Agencia Estatal de Investigación and the European Regional Development Fund (ERDF), as well as the ARTEMISA International Chair of Cybersecurity and the DANGER Strategic Project of Cybersecurity, both funded by the Spanish National Institute of Cybersecurity through the European Union -- NextGenerationEU and the Recovery, Transformation and Resilience Plan.

We would like to express our sincere gratitude to the anonymous reviewers for
their valuable feedback and suggestions that have improved the quality
of this work.

\bibliographystyle{elsarticle-num}
\bibliography{references}

\bio{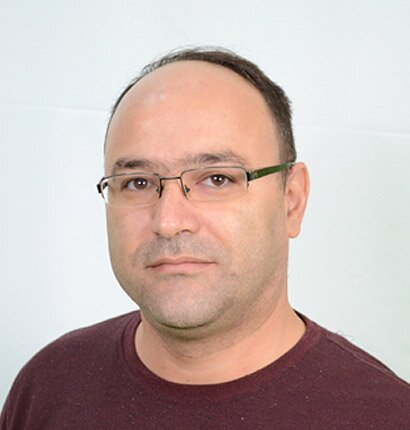}
{Dr. Hamza Kheddar} is an {Associate Professor} at the University of Medea and a Researcher at LSEA Lab. Medea (Algeria). He was awarded a Ph.D. in Telecommunication from USTHB University (Algeria), a Magister degree in Electronics (Spoken Communication) from USTHB University, and an Engineer degree in Telecommunication from ENSTTIC (Oran-Algeria). He obtained the Habilitation to Direct Research, which granted him the official authorization to supervise research, in June 2021.  His research interests include but are not limited to speech steganography, digital watermarking, data hiding, speech processing, intrusion detection, coding, and telecommunications. He is a reviewer for \textit{Computers \& Security}, \textit{IET Information Security}, \textit{IEEE Access}, \textit{KFUPM journals},  \textit{Applied Intelligence}, \textit{e-prime}, and \textit{Journal of image and graphic}.
\endbio

\bio{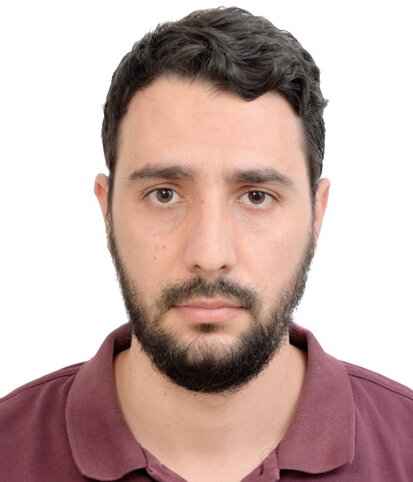}
{Dr. Mustapha Hemis} received the B.S. degree in Electrical Engineering and the M.Sc. degree in Telecommunication
and Multimedia and the PhD. degree in Telecommunication and Information Processing  from the University of Science and Technology Houari Boumediene (USTHB),
Algiers, Algeria, in 2010, 2012 and 2017, respectively. He is currently an Associate Professor 
 at the USTHB university and a researcher at LCPTS laboratory. His 
research interests include multimedia security, digital audio watermarking, steganography, steganalysis and data hiding. 

\endbio
\vspace{0.7cm}

\bio{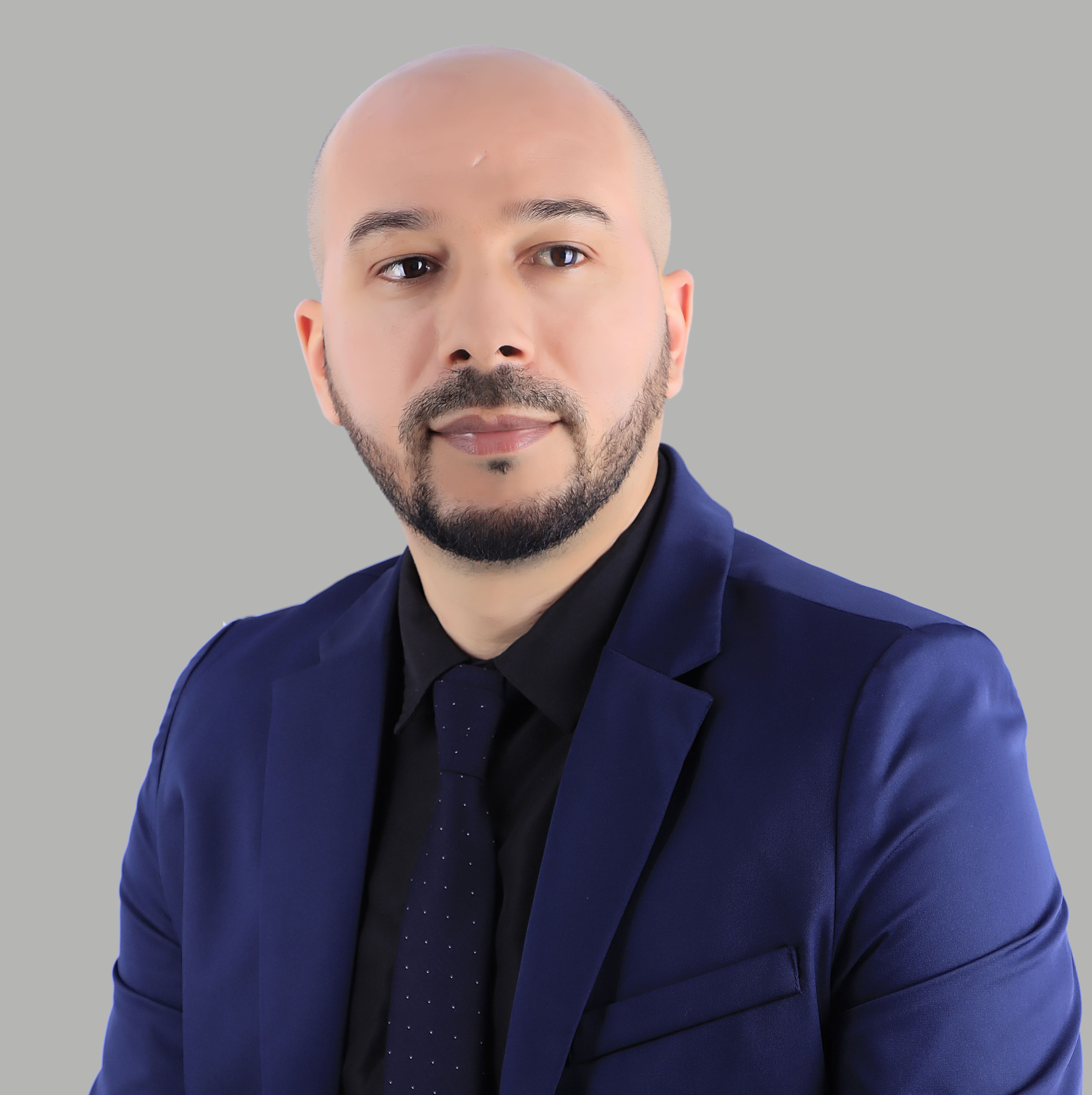}
{Dr. Yassine Himeur}  is presently an Assistant Professor of Engineering \& Information Technology at the
University of Dubai. He completed both his M.Sc. and Ph.D. degrees in Electrical Engineering in 2011 and 2015, respectively. Following his doctoral studies, he obtained the Habilitation to Direct Research, which granted him the official authorization to supervise research, in July 2017. His academic journey led him
to join the faculty at the University of Dubai after serving as a Postdoctoral Research Fellow at Qatar University from 2019 to 2022. Prior to that, he held the position of Senior Researcher from 2013 to 2019 at the Algerian Center for Development of Advanced Technologies (CDTA), where he also served as the
Head of the TELECOM Division from 2018 to 2019. Throughout his career, he has been actively involved in conducting R\&D projects and has played a significant role in proposing and co-leading several research proposals under the NPRP grant (QNRF, Qatar). With more than 100 research publications in
high-impact venues, he has made valuable contributions to the field. He was honored to receive the Best Paper Award at the 11th IEEE SIGMAP in Austria in 2014. His current research interests encompass Big Data and IoTs, AI/ML/DL, Healthcare Technologies, Digital Twins, Metaverse, Recommender
Systems, Smart grid, Building Energy Management, Smart Buildings, NLP, Generative AI, and
Cybersecurity.
\endbio

\bio{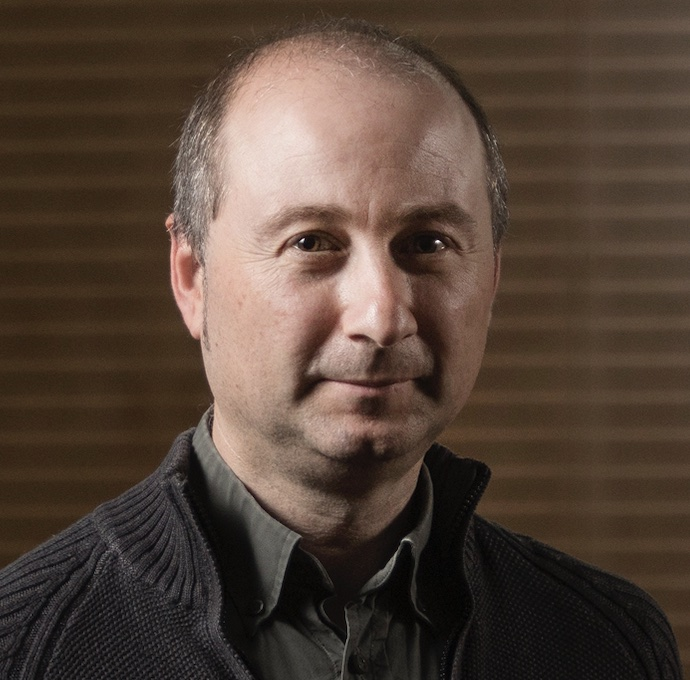}
{Prof. David Megías} is full Professor and the Principal Investigator of the KISON research group of the Internet Interdisciplinary Institute (IN3), at the Universitat Oberta de Catalunya (UOC). He received the Ph.D. degree in computer science from the Universitat Autònoma de Barcelona (UAB) in July 2000. Since October 2001, he has been at the UOC with a permanent position (currently as Professor). At the UOC, he has held several academic positions, until he was appointed director of the IN3 in April 2015. His current teaching is mostly related to computer networks, information security (watermarking and steganography), and research techniques and methodologies in the field of network and information technologies. His current research interests focus on information security and privacy, and include the security and privacy in multimedia content distribution (mainly in the watermarking and fingerprinting topics), steganography and steganalysis, and privacy concerns in different applications of decentralized networks. He has published more than 130 research papers in numerous international journals and conferences, 40 of them in journals indexed in JCR, and has participated in several national joint research projects both as a contributor and as principal investigator. He has supervised four doctoral theses and is a member of the editorial board and programme committees of several journals and conferences in the area of security and privacy.

\endbio

\bio{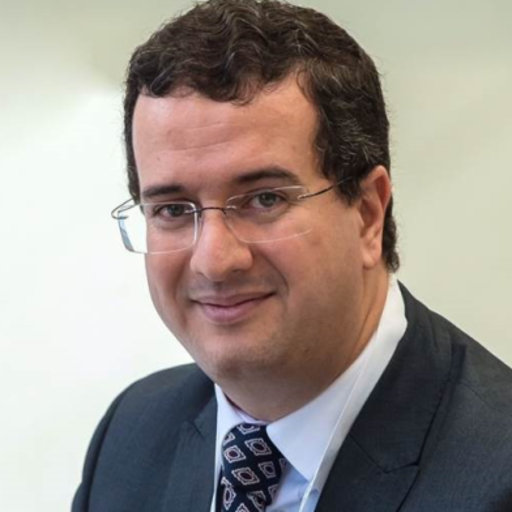}
{Prof. Abbes Amira} received his Ph.D degree from Queen’s University Belfast, United Kingdom and he has taken academic, leadership and consultancy positions in the United Kingdom including Queen’s University Belfast, Brunel University and De Montfort University in Leicester, France, Malaysia, and the Middleast at Qatar University, Qatar.  He is currently the Dean of the College of Computing and Informatics at the University of Sharjah, UAE.  During his career to date, Prof. Amira has been successful in securing substantial funding from government agencies and industry; he has supervised more than 25 PhD students and has over 350 publications in top journals and conferences in the area of embedded systems, artificial intelligence, IoT, image and signal processing. He obtained many international awards, including the 2008 VARIAN prize offered by the Swiss Society of Radiobiology and Medical Physics, CAST award, DELL-EM Envision the future (2018), IET premium award (2017) and many best paper and recognition awards in IEEE international conferences and events.

\endbio

\vspace{1cm}
\medskip

\end{document}